\documentclass[prb,twocolumn,showpacs,superscriptaddress,floatfix]{revtex4}
\usepackage{graphicx,amsfonts,amssymb,amsmath}
\usepackage{hyperref}

\newif\ifhyper
\hypertrue
\ifhyper
\hypersetup{
 citecolor = {green},
 colorlinks = {true},
 urlcolor = {blue} 
}
\fi

\newcommand{\n}{{\bf n}_1} 
\newcommand{\nn}{{\bf n}_2} 
\newcommand{\rmi}{\mathrm{i}}

\newcommand{\q}{{\bf q}} 
\newcommand{\ana}{a^{\phantom\dagger}}
\newcommand{\cra}{a^\dagger}
\newcommand{\anb}{b^{\phantom\dagger}}
\newcommand{\crb}{b^\dagger}
\newcommand{\bn}{{\boldsymbol{n}}}
\newcommand{\bi}{{\boldsymbol{i}}}
\newcommand{\bj}{{\boldsymbol{j}}}
\newcommand{\bk}{{\boldsymbol{k}}}
\newcommand{\bl}{{\boldsymbol{l}}}

\begin{document}

\title{Perturbative approach to an exactly solved problem~: the Kitaev honeycomb model}

\author{Julien Vidal}
\email{vidal@lptmc.jussieu.fr}
\affiliation{Laboratoire de Physique Th\'eorique de la Mati\`ere Condens\'ee,
  CNRS UMR 7600, Universit\'e Pierre et Marie Curie Paris 06, 4 Place Jussieu,
  75252 Paris Cedex 05, France}

\author{Kai Phillip Schmidt}
\email{schmidt@fkt.physik.uni-dortmund.de}
\affiliation{Lehrstuhl f\"ur theoretische Physik, Otto-Hahn-Stra\ss e 4, D-44221
  Dortmund, Germany}

\author{S\'ebastien Dusuel}
\email{sdusuel@gmail.com}
\affiliation{Lyc\'ee Saint-Louis, 44 Boulevard Saint-Michel, 75006 Paris, France
}
\begin{abstract} 

We analyze the gapped phase of the Kitaev honeycomb model perturbatively in the
isolated-dimer limit.
Our analysis is based on the continuous unitary transformations method which
allows one to compute the spectrum as well as matrix elements of operators
between eigenstates, at high order.
The starting point of our study consists in an exact mapping of the
original honeycomb spin system onto a square-lattice model involving an
effective spin and a hardcore boson.
We then derive the low-energy effective Hamiltonian up to order 10 which is
found to describe an interacting-anyon system, contrary to the order 4 result
which predicts a free theory.
These results give the ground-state energy in any vortex sector and thus also
the vortex gap, which is relevant for experiments.
Furthermore, we show that the elementary excitations are emerging free fermions
composed of a hardcore boson with  an attached spin- and phase- operator string.
We also focus on observables and compute, in particular,  the spin-spin correlation functions. We show that they admit a multi-plaquette expansion that we derive up to order 6. 
Finally, we study the creation and manipulation of anyons with local operators, show that they also create fermions, and discuss the relevance of our findings for experiments in optical
lattices.

\end{abstract}


\maketitle

%
%
\section{Introduction}
\label{sec:intro}
%
%

Elementary particles can be classified in two categories according to the value of their spin. 
Half-integer spin particles obey Fermi-Dirac statistics and are called fermions whereas integer-spin particles obey Bose-Einstein statistics and are known as bosons.
However, some quantum objects may obey other (fractional) statistics describing nontrivial braiding as initially suggested by Leinaas and Myrheim more than thirty years ago \cite{Leinaas77}, and by Wilczek in the eighties \cite{Wilczek82_1, Wilczek82_3}. 
Despite numerous theoretical works, these so-called anyons are still waiting for a direct observation although recent experimental proposals are very promising  (see Ref.~\onlinecite{Feldman07}).

In the last years, anyons have drawn much attention because of their interest for topological quantum computation \cite{Preskill_HP}.  
In this perspective, several models have been proposed among which the celebrated toric code \cite{Kitaev03} which is a spin-1/2 system whose elementary excitations behave as semions. 
However, the experimental realization of this system is rather tricky since it involves four-spin interactions. 
Here, we shall focus on another system originally proposed by Kitaev \cite{Kitaev06} which only involves two-spin interactions. 
This model is very rich since it contains Abelian and non-Abelian anyonic as well as fermionic excitations. 
Thus, it has been the subject of many recent studies concerning the spectrum,
\cite{Pachos07,Feng07,Lee07,Chen07_1,Chen08,Schmidt08,Lahtinen08,Kells08,Dusuel08_1}
the correlation functions and the entanglement \cite{Dusuel08_1,Baskaran07,Yang08,Zhao08,Gu08}, or the quench dynamics \cite{Sengupta08,Mondal08}. 
Let us also mention several extensions \cite{Yu07,Yang07,Mandal08} among which the analysis of time-reversal symmetry breaking terms \cite{Yao07,Dusuel08_2} which may give rise to a chiral spin liquid.

Furthermore, this model is susceptible to be realized in various experimental systems such as polar molecules, ultracold atoms \cite{Duan03,Micheli06,Jiang08,Aguado08} or Josephson junctions \cite{Nori08}. It thus constitutes an appealing candidate for the observation of anyons. 
Nevertheless, the presence of fermions in the spectrum may spoil the detection process~; a point completely missed in a recent proposal (see Ref.~\onlinecite{Vidal08_1} for explanation and Ref.~\onlinecite{Dusuel08_1} for details).

The goal of the present paper is to investigate the gapped phase of the Kitaev honeycomb model \cite{Kitaev06}. Indeed, in his remarkable seminal paper, Kitaev mainly focuses on the special subspace of the Hilbert space to which the ground state belongs to and the low-energy spectrum of other subspaces has only been discussed lately \cite{Schmidt08}. 
Our aim is to bridge this gap by providing a high-order perturbative analysis, in the isolated-dimer limit, of the spectrum as well as some interesting results about the creation and the manipulation of anyons which is of relevance for experiments \cite{Jiang08,Aguado08}.
Part of our results have already been given in two short papers \cite{Schmidt08,Dusuel08_1} and the present paper may be considered as an extended and detailed version of these works. However many other results are presented here among which the interplay between fermions and anyons under string operations discussed in Sec.~\ref{sec:Observables}.  

This paper is organized as follows. In the next section, we introduce the model as well as its main properties. In particular, we discuss the importance of the boundary conditions and insist on the role played by conserved quantities \cite{Kells08} and the constraints resulting from them.
In Sec.~\ref{sec:mapping}, we show how to map the Kitaev model involving spins on the honeycomb lattice onto an effective spin and hardcore boson on a square lattice. This mapping is the starting point of the perturbation theory presented in this work. 
In Sec.~\ref{sec:perturbation}, we explain how to diagonalize the Hamiltonian order by order using the perturbative continuous unitary transformation (PCUT) method . 
The study of the low-energy (zero-quasiparticle) sector is the subject of Sec.~\ref{sec:zeroqp}, a large part of which is devoted to a pictural (and hopefully pedagogical) analysis and construction of the
eigenstates of the toric code model which naturally emerges from this problem. There, we give the perturbative expansion form of the ground state energy for any vortex configuration. The effective low-energy theory is found to be described by interacting anyons contrary to the lowest-order result which predicts free anyons \cite{Kitaev06}.
Sec.~\ref{sec:oneqp} focuses on the study of the one-quasiparticle subspace, where the physics is shown to be that of a particle hopping in a magnetic field with zero or half a flux quantum per elementary plaquette. 
The demonstration of the fermionic nature (known from exact solutions) of the quasiparticles is briefly sketched in Sec.~\ref{sec:multiqps}. 
In Sec.~\ref{sec:checks}, we provide some checks of our results by analyzing simple vortex configurations which allow for an exact solution. 
The spin-spin correlation functions and the manipulation of anyons are tackled in 
Sec.~\ref{sec:Observables}, which is devoted to the renormalization of observables. 
Finally, we discuss several issues and give some perspectives.
Technical details as well as all relevant coefficients involved in the perturbative expansions are gathered in appendices. 

In what follows, we tried to be as pedagogical as possible and always favored simple demonstrations on concrete examples rather than lengthy proofs for general situations. We hope that it will help the reader to understand the richness of this model. 

%
%
\section{The model}
\label{sec:model}
%
%

\subsection{Hamiltonian and boundary conditions}

\begin{figure}[t]
  \includegraphics[width=\columnwidth]{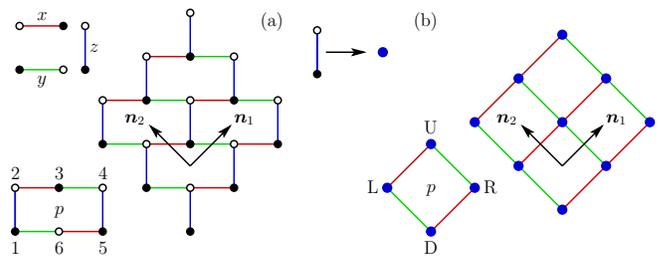}
  \caption{(color online).
    Mapping of the honeycomb (brick-wall) lattice onto a square
    lattice with unit basis vectors $\bn_1$ and $\bn_2$. Each $z-$dimer with
    four spin configurations is replaced by a single site with four degrees of
    freedom~: the occupation number of hardcore boson (0 or 1), and the
    effective spin ($\Uparrow$, or $\Downarrow$) which is chosen as the spin
    of the black site of the considered $z-$ dimer. The numbering of the sites
    of a plaquette $p$ is shown in both cases.}
  \label{fig:mapping_brickwall_square}
\end{figure}

The model considered in this work is a spin-$1/2$ system proposed by Kitaev
\cite{Kitaev06} in which spins are located at the vertices of a honeycomb
lattice. Since the honeycomb lattice is topologically equivalent to the
brick-wall lattice,  we shall always represent it as shown in
Fig.~\ref{fig:mapping_brickwall_square}a.
In this lattice, one distinguishes three types of links ($x$, $y$, and $z$) to
which one associates three different couplings and interactions.
The Hamiltonian of the system is
%
%
\begin{equation}
  \label{eq:ham}
  H=-\sum_{\alpha=x,y,z}\sum_{\alpha-\mathrm{links}}
  J_\alpha\,\sigma_i^\alpha\sigma_j^\alpha,
\end{equation} 
%
%
where $\sigma_i^\alpha$ are the usual Pauli matrices at site $i$.
In the following we assume, without loss of generality\cite{Kitaev06}, that
$J_\alpha \geq 0$ for all $\alpha$ and $J_z\geq J_x,J_y$.

\begin{figure}[t]
  \includegraphics[width=\columnwidth]{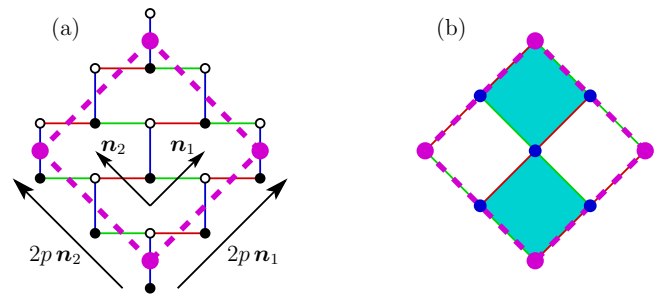}
  \caption{(color online).
    The periodic boundary conditions used in this work, (a) on the original
    brick-wall lattice and (b) on the effective square lattice of $z$-dimers
    (see Fig.~\ref{fig:mapping_brickwall_square}). In the figure $p=1$.
    In both cases, the finite-size system is put on a torus obtained by
    identifying the opposite sides of the dashed (magenta) square. For clarity,
    the site at the point chosen as the origin has been depicted bigger (and in
    magenta). Half the square plaquettes in (b) have been colored in cyan (gray) to show
    that the periodic boundary conditions allow us to bi-color the lattice.}
  \label{fig:pbc}
\end{figure}

We will either work with an infinite system and open boundary conditions (a
plane), or with a finite (or infinite) system and periodic boundary conditions
(a torus). In the latter case and for reasons that will become clearer in the
following (in particular, see Sec.~\ref{sec:sub:toricode}), we shall restrict
ourselves to the periodic boundary conditions (PBC) depicted in
Fig.~\ref{fig:pbc}.
The number of sites $N_\mathrm{s}$ is $N_\mathrm{s}=2(2p)^2=8p^2$, with
$p\in\mathbb{N}$ ($p=1$ in Fig.~\ref{fig:pbc}a).
Let us anticipate what follows and mention that these boundary conditions are
such that the lattice of $z$-dimers (Fig.~\ref{fig:mapping_brickwall_square}b)
can be bi-colored as shown in Fig.~\ref{fig:pbc}b).
%
%
%
%

%
%
%
%

\subsection{Conserved quantities}
\label{sec:sub:model_conserved_quantities}

\begin{figure}[t]
  \includegraphics[width=0.6\columnwidth]{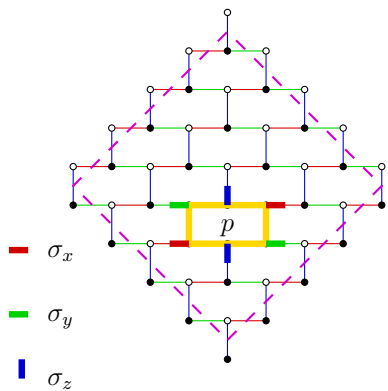}
  \caption{(color online). Illustration of the conserved plaquette quantity
    $W_p$. The thick yellow (lightest gray) line delimitates the plaquette $p$. The thick red (gray), green (light gray)
    or blue (dark gray) segments represent the Pauli matrices $\sigma_i^{\mathrm{out}(i)}$.}
  \label{fig:Wp}
\end{figure}

A remarkable property of Hamiltonian (\ref{eq:ham}), is that its elementary
operators $K_{ij}=\sigma_i^\alpha\sigma_j^\alpha$ commute with plaquette
operators $W_p$ so that $\left[H,W_p\right]=0$. For the plaquette $p$ shown in Fig.~\ref{fig:mapping_brickwall_square}a, such an operator is defined as
%
%
\begin{equation}
    \label{eq:Wp}
    W_p=K_{12}K_{23}K_{34}K_{45}K_{56}K_{61}
    =\sigma_1^x \sigma_2^y \sigma_3^z \sigma_4^x \sigma_5^y \sigma_6^z.
\end{equation}
%
%
Let us mention that in the expression of $W_p$ in terms of the $K$'s, one could
have started at any site instead of site $1$ and/or one could have taken the
product of $K's$ anti-clockwise instead of clockwise. Furthermore, the
expression in terms of $\sigma$'s could also be written
$W_p=\prod_i\sigma_i^{\mathrm{out}(i)}$ where $i$ runs over the set of six
spins around the plaquette $p$, and where the notation $\mathrm{out}(i)$ means
the ``outgoing'' direction at site $i$, with respect to the plaquette's contour.
An illustration of the $W_p$ operator is given in Fig.~\ref{fig:Wp}.

%
%
%
%

Since $W_p^2=\mathbb{I}$, the eigenvalues of the plaquette operators are
$w_p=\pm 1$. Note that $\left[W_p,W_{p'}\right]=0$, as can be shown from the usual
Pauli matrices algebra.
As a consequence, $H$ and the $W_p$'s can be diagonalized simultaneously.
Following Kitaev, we will call a vortex sector a subspace of the Hilbert space
with a given map of the $w_p$'s. By definition a vortex is
a plaquette for which $w_p=-1$, so that for example, the vortex-free sector is
defined by $w_p=+1$ for all $p$'s.

\begin{figure}[t]
  \includegraphics[width=\columnwidth]{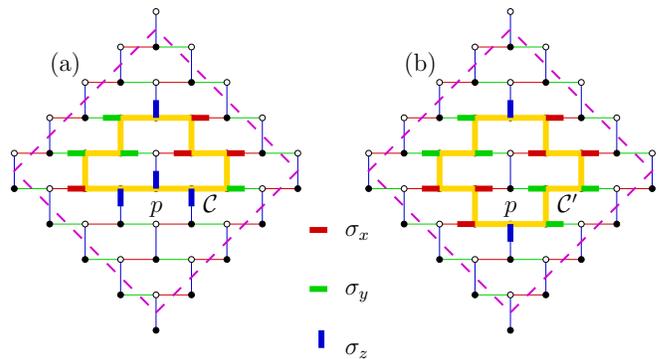}
  \caption{(color online). Illustration of the relation
    $\mathcal{L}'=W_p\mathcal{L}$, with
    $\mathcal{L}=\prod_{i\in\mathcal{C}}\sigma_i^{\mathrm{out}(i)}$, $W_p$ 
    already shown in Fig.~\ref{fig:Wp} and
    $\mathcal{L}'=\prod_{i\in\mathcal{C'}}\sigma_i^{\mathrm{out}(i)}$.
    The thick yellow (lightest gray) line in (a) represents the contour $\mathcal{C}$ and the
    one in (b) represents $\mathcal{C}'$. As in Fig.~\ref{fig:Wp}, the thick
    red (gray), green (light gray) or blue (dark gray) segments represent the Pauli matrices
    $\sigma_i^{\mathrm{out}(i)}$.}
  \label{fig:W_C}
\end{figure}

\begin{figure}[t]
  \includegraphics[width=\columnwidth]{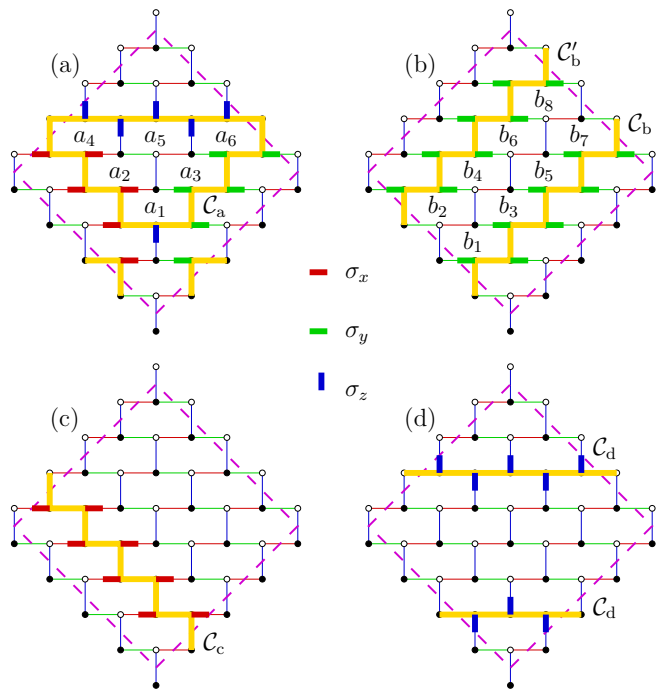}
  \caption{(color online). Examples of conserved loop operators
    $\mathcal{L}=\prod_{i\in\mathcal{C}}\sigma_i^{\mathrm{out}(i)}$
    in a finite-size system with PBC.}
  \label{fig:loop_operators}
\end{figure}

In fact, all loop operators made of ``outgoing spins'' (see
Figs.~\ref{fig:W_C} and \ref{fig:loop_operators}) are conserved and all commute
with each other, which can be verified in the same way as for the $W_p$'s.
However, not all of them can be set independently to $\pm 1$. Some relations
among them arise from the following fact (which can be checked by studying all
possible cases)~: the product of $W_p$ and a nearby loop operator $\mathcal{L}$
gives a new loop operator $\mathcal{L}'=W_p \mathcal{L}$, as illustrated on a
particular example in Fig.~\ref{fig:W_C}.

%
%
%
%

As an illustration of other relations involving loop operators around the torus,
with the loops of Fig.~\ref{fig:loop_operators} one has
%
%
\begin{eqnarray}
  &&\mathcal{L}_\mathrm{a}=\prod_{n=1}^6W_{a_n},\\
  &&\mathcal{L}_\mathrm{b}'=\mathcal{L}_\mathrm{b}\prod_{n=1}^8W_{b_n},\\
  \label{eq:WdCbCc}
  &&\mathcal{L}_\mathrm{d}=-\mathcal{L}_\mathrm{a}\mathcal{L}_\mathrm{b}
  \mathcal{L}_\mathrm{c},
\end{eqnarray}
%
%
where we have denoted, for example,
$\mathcal{L}_\mathrm{a}=\prod_{i\in\mathcal{C}_\mathrm{a}}
\sigma_i^{\mathrm{out}(i)}$. The minus sign in the last equation above
comes from the crossing of $\mathcal{L}_\mathrm{b}$ and
$\mathcal{L}_\mathrm{c}$.
In the three expressions above, the product of plaquette operators
could also have been taken over the complementary set of plaquettes. Indeed,
on the torus the relations among loop operators yield the following constraint
%
%
\begin{equation}
  \label{eq:prodW}
  \prod_{\mathrm{all}\; p\mathrm{'s}}W_p=\mathbb{I},
\end{equation}
%
%
showing in particular that {\em the number of vortices has to be even in a
  system with PBC}.

%
%
%
%

From examples shown in Fig.~\ref{fig:loop_operators}, one can deduce that
all $W_p$'s [except one, because of Eq.~(\ref{eq:prodW})],
$\mathcal{L}_\mathrm{b}$ and $\mathcal{L}_\mathrm{c}$ can be set independently
to $\pm 1$, which then imposes all other conserved quantities. 

\subsection{Some results from the exact solution}

The above discussed local conserved quantities are not sufficient to fully
diagonalize the Hamiltonian. Indeed, if $N_\mathrm{s}$ is the number of sites,
then there is a total of $N=N_\mathrm{s}/2$ plaquettes, but only $N-1$
independent ones. With the two cycles around the torus, this gives $N+1$
independent conserved quantities, which is obviously smaller than
$N_\mathrm{s}$.

However, $H$ has a crucial property~: it can be transformed into a free
Majorana fermions Hamiltonian and is thus exactly solvable. Let us also mention
that another solution based on the Jordan-Wigner transformation
maps the spin Hamiltonian $H$ onto a spinless
fermions system with $p$-wave pairing \cite{Feng07,Chen07_1,Chen08}.

As shown by Kitaev \cite{Kitaev06}, the ground state of $H$ lies in the vortex-free sector and the phase diagram contains,  in this sector,  two phases~:
a gapped phase for $J_z> J_x+J_y$ and a gapless phase for
$J_z \leq J_x+J_y$. In the gapped phase the low-energy excitations are Abelian
anyons (semions) whereas in the gapless phase, the low-energy excitations are
fermionic. The gapless phase acquires a gap in the presence of a
magnetic field and then contains gapped non-Abelian anyon excitations.
The phase diagram has also been investigated in other vortex configurations
such as the vortex-full sector and similar phases have been obtained.
More precisely, one has a gapped phase for $J_z^2> J_x^2+J_y^2$ and a gapless
phase in the opposite case\cite{Lahtinen08}.

Our goal here is to determine the low-energy spectrum for any vortex configuration. Of course, one may use the fermionic Hamiltonian mentioned above but, it can only be exactly diagonalized for translation-invariant configuration. 
Here, we follow an alternative route by focusing on the isolated-dimer limit $J_z \gg J_x,J_y$.

%
%
\section{Mapping onto an effective spin boson problem}
\label{sec:mapping}
%
%

\subsection{Mapping of the Hamiltonian}

The very first step of our analysis consists in mapping the four possible
states of the two spins of a $z$-dimer onto those of an effective spin and a
hardcore boson.
More precisely, denoting $|\!\uparrow \rangle$ ($|\!\downarrow \rangle$) the
eigenstate of $\sigma_z$ with eigenvalue $+1$ ($-1$), an isolated $z$-dimer
can be in one of the two low-energy states
$\{|\!\uparrow\uparrow\rangle,|\!\downarrow\downarrow\rangle\}$ with energy
$-J_z$, or in one of the two high-energy states
$\{|\!\uparrow\downarrow\rangle,|\!\downarrow\uparrow\rangle\}$ with energy
$+J_z$.
Keeping in mind that our aim is to perform a perturbation theory in the limit
$J_z \gg J_x,J_y$, it is natural to interpret the change from a ferromagnetic to
an anti-ferromagnetic configuration as the creation of a particle with an
energy cost $2 \, J_z$.
By construction, such a particle is a hardcore boson. The remaining degree of
freedom can be described by a spin 1/2, indicating which of the two
configurations is realized.
There are many possible parametrizations but here we choose the following
%
%
\begin{equation}
  \label{eq:mappings}
  |\!\uparrow\uparrow\rangle=|\!\Uparrow\! 0\rangle,
  |\!\downarrow\downarrow\rangle=|\!\Downarrow\! 0\rangle,\,
  |\!\uparrow\downarrow\rangle=|\!\Uparrow\! 1\rangle,
  |\!\downarrow\uparrow\rangle=|\!\Downarrow\! 1\rangle.
\end{equation}
%
%
The left (right) spin is the one of the black (white) site of the dimer
($|\!\uparrow\downarrow\rangle=|\!\uparrow_\bullet\downarrow_\circ\rangle$,
etc). Double arrows represent the state of the effective spin which is here the
same as the state of the left  (black) spin.

Within such a mapping, effective spins and hardcore bosons live on the
effective square lattice of $z$-dimers (see
Fig.~\ref{fig:mapping_brickwall_square}). This lattice is shown again
in Fig.~\ref{fig:pbc}b, together with the PBC, which are such that it can be bi-colored. In what follows, the sites of the effective lattice will be denoted with bold letters, such as $\bi$.

Let us now write the Hamiltonian (\ref{eq:ham}) in this language. Therefore, we first translate the action of the spin operators in the
effective-spin boson (ESB) formalism. It is easy to check that one has
%
%
\begin{equation} \label{eq:mapping}
\begin{array}{lcl}
 \sigma_{\bi,\bullet}^x=\tau_\bi^x (\crb_\bi+\anb_\bi)
  &,&
 \sigma_{\bi,\circ}^x=\crb_\bi+\anb_\bi ,\\ 
 \sigma_{\bi,\bullet}^y=\tau_\bi^y (\crb_\bi+\anb_\bi)
  &,&
  \sigma_{\bi,\circ}^y=\mathrm{i} \, \tau^z_\bi (\crb_\bi-\anb_\bi) , \\
  \sigma_{\bi,\bullet}^z=\tau_\bi^z
  &,&
  \sigma_{\bi,\circ}^z=\tau_\bi^z (1-2 \crb_\bi \anb_\bi).
\end{array}
\end{equation}
%
%
The operators $\tau^\alpha_\bi$ ($\alpha=x,y,z$) are the Pauli matrices
acting on the effective spin at site $\bi$, while $\anb_\bi$ and $\crb_\bi$ are
hardcore bosonic annihilation and creation operators, satisfying the usual
on-site anticommutation relation $\{\anb_\bi,\crb_\bi\}=\mathbb{I}$
(and operators on different sites commute).
Setting once for all $J_z=1/2$ so that creating a boson costs an energy $1$
in the isolated-dimer limit, the Hamiltonian (\ref{eq:ham}) reads 
%
%
\begin{equation}  \label{eq:ham_v2}
  H = -\frac{N}{2}+Q+T_0+T_{+2} + T_{-2},
 \end{equation}
%
%
where $N$ is the number of $z$-dimers (or, equivalently, of square plaquettes), and 
%
%
\begin{eqnarray}
  Q&=&\sum_\bi\crb_\bi\anb_\bi , \label{eq:defQ}\\
  T_0&=&-\sum_\bi\left(J_x\, t_\bi^{\bi+\bn_1}+J_y\, t_\bi^{\bi+\bn_2}
    +\mathrm{h.c.}\right),\\
  T_{+2}&=&-\sum_\bi\left(J_x\, v_\bi^{\bi+\bn_1}+J_y\, v_\bi^{\bi+\bn_2}\right)
  =(T_{-2})^\dagger.
\end{eqnarray}
%
%
These operators are built from local hopping and pair creation operators
%
%
\begin{equation}
\label{eq:deftv}
\begin{array}{lcl}
  t_\bi^{\bi+\bn_1}=\crb_{\bi+\bn_1}\anb_\bi\, \tau^x_{\bi+\bn_1}
  &,&
  t_\bi^{\bi+\bn_2}=-\rmi\, \crb_{\bi+\bn_2}\anb_\bi\,
  \tau^y_{\bi+\bn_2}\tau^z_\bi , \\
  v_\bi^{\bi+\bn_1}=\crb_{\bi+\bn_1}\crb_\bi\, \tau^x_{\bi+\bn_1}
   &,& 
  v_\bi^{\bi+\bn_2}=\rmi\, \crb_{\bi+\bn_2}\crb_\bi\, \tau^y_{\bi+\bn_2}\tau^z_\bi.
\end{array}
\end{equation}
%
%
We emphasize that the mapping (\ref{eq:mapping}) explicitly breaks the symmetry
between white and black sites of the original brick-wall lattice. This is
responsible for the apparent breaking of symmetry between the $x/\bn_1$ and
$y/\bn_2$ directions in Eq.~(\ref{eq:deftv}). However, for all the physically
observable results, this symmetry remains intact (see for example the series
expansion of eigenenergies in Appendix~\ref{app:zeroqp}). Note however that the
$\bn_1+\bn_2$ and $\bn_1-\bn_2$ directions are not equivalent, as can be seen from the
underlying brick-wall lattice.

\subsection{Conserved quantities}

\begin{figure}[t]
    \includegraphics[width=\columnwidth]{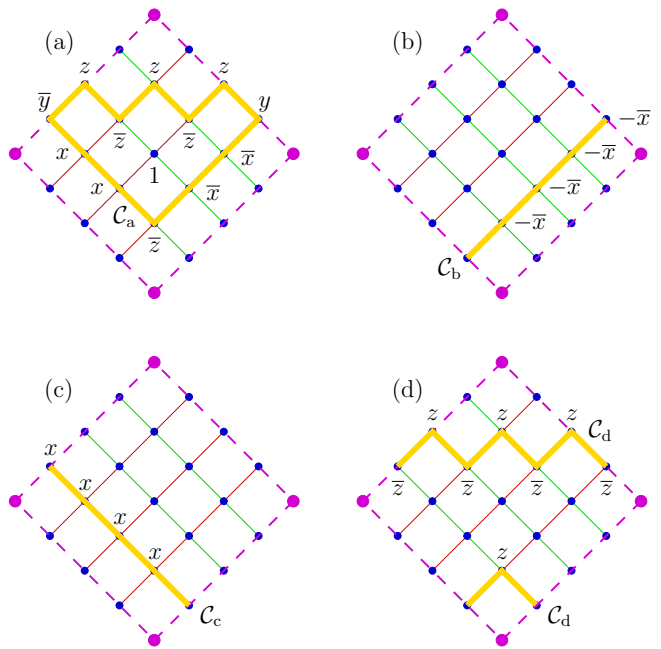}
    \caption{(color online). On the four figures, the thick yellow (lightest gray) line
    represents the contours $\mathcal{C}$. 
    The operators $\omega_\bi$ which are such that the loop operators read
    $\mathcal{L}=\prod_{\bi\in\mathcal{C}}\omega_\bi$
    are indicated in the figures. In the present case, they can take the following values
    $x=\tau^x$, $\overline{x}=(-1)^{b^\dagger b}\tau^x$ (and the same for $y$
    and $z$), and $-\overline{x}=-(-1)^{b^\dagger b}\tau^x$. These figures 
    are the same as the ones of Fig.~\ref{fig:loop_operators}, but on the
    effective lattice.}
    \label{fig:loop_operators_square}
\end{figure}

Let us now rephrase the conserved operators discussed in Sec.~\ref{sec:sub:model_conserved_quantities} in the effective language.
Using the notations depicted in Fig.~\ref{fig:mapping_brickwall_square}b, as
well as the mapping (\ref{eq:mapping}), the plaquette operators transform into
%
%
\begin{equation}
    \label{eq:Wp_eff}
    W_p=(-1)^{\crb_{\mbox{\tiny L}}\anb_{\mbox{\tiny L}}
    +\crb_{\mbox{\tiny D}}\anb_{\mbox{\tiny D}}}\,
    \tau_{\mbox{\tiny L}}^y\,\tau_{\mbox{\tiny U}}^z\,
    \tau_{\mbox{\tiny R}}^y\,\tau_{\mbox{\tiny D}}^z.
\end{equation}
%
%
Note that $(-1)^{\crb_\bi\anb_\bi}=1-2\crb_\bi\anb_\bi$. In the same vein,
for the cycles around the torus shown in Figs.~\ref{fig:loop_operators}b-\ref{fig:loop_operators}c, which are reproduced for the effective lattice in
Figs.~\ref{fig:loop_operators_square}b-\ref{fig:loop_operators_square}c,
one has
$\mathcal{L}_\mathrm{b}=\prod_{\bi\in\mathcal{C}_\mathrm{b}}
\left[-(-1)^{\crb_\bi\anb_\bi}\tau^x_\bi\right]
=\prod_{\bi\in\mathcal{C}_\mathrm{b}}
\left[(-1)^{\crb_\bi\anb_\bi}\tau^x_\bi\right]$ (since there is an even number
of sites on the contour with the PBC chosen here), as well as
$\mathcal{L}_\mathrm{c}=\prod_{\bi\in\mathcal{C}_\mathrm{c}}\tau^x_\bi$.
The expression for
$\mathcal{L}_\mathrm{d}$ (see Fig.~\ref{fig:loop_operators}d), namely
$\mathcal{L}_\mathrm{d}=\prod_{\bi\in\mathcal{C}_\mathrm{d}}\omega_\bi$, is a
bit more complicated, but it should be clear from
Fig.~\ref{fig:loop_operators_square}d what the $\omega_\bi$'s are.
Finally, for the contour shown in Fig.~\ref{fig:loop_operators_square}a
(which is in correspondence with Fig.~\ref{fig:loop_operators}a), one has
$\prod_{p\subset\mathcal{C}_\mathrm{a}}W_p
=\prod_{\bi\in\mathcal{C}_\mathrm{a}}\omega_\bi=\mathcal{L}_\mathrm{a}$ with
the $\omega_\bi$'s indicated in the figure, and with
$p\subset\mathcal{C}_\mathrm{a}$ meaning the plaquettes $p$ enclosed in the
contour $\mathcal{C}_\mathrm{a}$. With these notations, one can easily check
that Eq.~(\ref{eq:WdCbCc}) still holds.
%
%
%
%

The elementary hopping and pair creation operators, namely $t_\bi^\bj$ and
$v_\bi^\bj$ with $\bi$ and $\bj$ nearest neighbors, have a very remarkable
property : they all commute with the $W_p$'s, as well as with any other loop
operator
%
%
\begin{equation}
  [t_\bi^\bj,W_p]=[v_\bi^\bj,W_p]=[t_\bi^\bj,\mathcal{L}]
  =[v_\bi^\bj,\mathcal{L}]=0.
\end{equation}
%
%

The original spin problem on the honeycomb lattice is thus mapped onto a
quadratic hardcore boson problem on an effective square lattice, with conserved
plaquette and loop operators. Let us underline that this mapping is exact and
just provides an alternative description of the spin problem. 
The resulting Hamiltonian (\ref{eq:ham_v2}) remains difficult to
diagonalize (except, of course, if one remembers that the model can be fermionized), since
({\it i}) bosons are hard core which prevents the use of a Bogoliubov
transformation, ({\it ii}) bosonic and spin degrees of freedom are correlated.
The conserved plaquette operators will of course be useful in simplifying
and solving the problem as recently underlined in Ref.~\onlinecite{Kells08}.

%
%
\section{Perturbation theory in the gapped phase}
\label{sec:perturbation}
%
%

\subsection{Effective Hamiltonian from PCUTs}

The starting point of the present perturbation theory is the isolated-dimer
limit, namely $J_x=J_y=0$. In this limit, the spectrum is made of equidistant
and degenerate levels separated by an energy gap $\Delta=2J_z=1$.
To compute the perturbative spectrum, there are of course several methods among
which the Green's function formalism initially used by Kitaev \cite{Kitaev06}. However,
if this approach is efficient to obtain the first nontrivial (nonconstant)
correction,  it becomes tricky to implement at higher orders.

Here, following Ref.~\onlinecite{Schmidt08}, we use an
alternative approach based on continuous unitary transformations (CUTs)
conjointly proposed by Wegner \cite{Wegner94} and G{\l}azek and Wilson
\cite{Glazek93,Glazek94}. We refer the interested reader to
Ref.~\onlinecite{Dusuel04_2} for a recent pedagogical introduction.
Its perturbative version denoted PCUTs  is
especially well-suited to the problem at hand. This technique is detailed in
several works \cite{Stein97,Knetter00}. 
Let us simply mention that the CUTs
method requires the choice of a generator that drives the flow of the operators.
All the  results given here have been obtained with the so-called quasiparticle number-conserving generator first proposed by Mielke \cite{Mielke98} for finite
matrices and generalized to many-body systems by Knetter and Uhrig\cite{Knetter00}.

The latter have computed the perturbative expansion for {\it any}
Hamiltonian of the form
%
%
\begin{equation}  \label{eq:hamCUT}
 H = Q+ T_{-2} +T_{-1}+T_0+T_{+1} +T_{+2},
 \end{equation}
%
%
provided two hypothesis are satisfied:
%
%
\begin{itemize}
    \item the unperturbed Hamiltonian $Q$ has an equidistant spectrum bounded
    from below~;
    \item the perturbing Hamiltonian $\sum_{n=-2}^{+2} T_n$ is such that
    $[Q,T_n]=n\, T_n$.
\end{itemize}
%
%
Clearly, the Hamiltonian (\ref{eq:ham_v2}) meets these two criteria (up to a
constant term) noting that in the present case, one has $T_{\pm 1}=0$.  
Here, we have included the "small" parameters, namely $J_x$ and $J_y$, in the
definition of the $T_n$ operators, which is not the convention usually adopted
in the CUTs community.

The CUTs method together with the quasiparticle number-conserving generator
unitarily transforms the Hamiltonian (\ref{eq:hamCUT}) into  an
effective Hamiltonian $H_\mathrm{eff}=U^\dagger H U$ commuting with $Q$,
$U$ being a unitary operator. 
We give the first terms of the expansion up to order 4 in Appendix \ref{app:PCUT}.
As can be seen in Table \ref{tab:Heff}, the number of terms appearing in the perturbative expansion quickly increases with the order.
For instance, at order 2, the effective Hamiltonian reads in our case
%
%
\begin{equation} \label{eq:heff}
  H_\mathrm{eff} = -\frac{N}{2}+Q+T_0-\frac{1}{2} T_{-2} T_{+2}
  +\frac{1}{2} T_{+2} T_{-2},
 \end{equation}
%
%
whereas at order 10, there are more than $10^4$ operators to consider.

Writing $H_\mathrm{eff}$ this way is only the very first part of the job since
one next has to ({\it i}) determine its action in each subspace of a given
quasiparticle (QP) number $q$~; ({\it ii}) diagonalize $H_\mathrm{eff}$ in each
of these subspaces. This is the object of the next sections~: we first
study the lowest-energy states ($q=0$ QP) which is the main contribution of
our work~; then we turn to the $q=1$ QP states and recover the high-energy gap
from the QP dispersion~;
we end by $q\geqslant 2$ QP states, whose properties determine the
statistics of the QPs, and we will see that the QPs behave as fermions which
are, furthermore, non interacting. 
This fact is at the origin of a tremendous simplification of the effective Hamiltonian. Indeed, we found that $H_\mathrm{eff}$ can be written, at all orders and in the thermodynamical limit, as
%
%
\begin{eqnarray}
    \label{eq:heff_explicit}
    H_\mathrm{eff} &=& E_0 + \mu \, Q
    - \sum_{\{p_1,\ldots,p_n\}} C_{p_1,\ldots,p_n}
    W_{p_1}\ldots W_{p_n}\qquad\\
    &&-\sum_{\{\bj_1,\ldots,\bj_n\}}
    D_{\bj_1,\ldots,\bj_n}
    S_{\bj_1,\ldots,\bj_n}\crb_{\bj_n}\anb_{\bj_1}.\nonumber
\end{eqnarray}
%
%
We shall discuss each term in detail in the following sections, but let us
mention that $E_0$, $\mu$, the $C$'s and the $D$'s are coefficients whose series expansion are computed. 
The $S$ operators are strings of spin operators $\tau_\bj^\alpha$ and of phase
factors $(-1)^{\crb_\bj\anb_\bj}$ on the cluster $\{\bj_1,\ldots,\bj_n\}$.
This very special form of multi-particle terms
[remember $(-1)^{\crb_\bj\anb_\bj}=1-2\crb_\bj\anb_\bj$], leading to phase
factors and spin-strings only, is responsible for the emergence of fermions in
the model.

For a finite-size system with PBC, new terms appear in the
effective Hamiltonian. They involve loop operators around the torus, and
appear at a minimal order being the linear size $2p$ of the lattice. Such loop
operators are associated to contours as the ones shown in
Figs.~\ref{fig:loop_operators} and \ref{fig:loop_operators_square},
namely $\mathcal{L}_\mathrm{b}$, $\mathcal{L}_\mathrm{c}$ and
$\mathcal{L}_\mathrm{b}'$ for the contours $\mathcal{C}_\mathrm{b}$,
$\mathcal{C}_\mathrm{c}$ and $\mathcal{C}_\mathrm{b}'$.
The presence of such loop operators in the effective Hamiltonian shows that the
eigenstates of the Hamiltonian are also eigenstates of these loop operators.
Their effect is to lift the degeneracies between states (which for each energy
is at least four in the thermodynamical limit, since some of the excitations are
Abelian semions and the genus of a torus is 1, see
Ref.~\onlinecite{Preskill_HP}). We shall not dive into the details of such
finite-size corrections, since our approach allows us to directly tackle with the
most interesting thermodynamical limit. However, let us make a remark about a
numerical check of this statement for small system sizes. For a torus whose
linear size is strictly smaller than 4, the loop operator terms around the
torus dominate the expansion over the $W_p$'s, and for a size of 4 both types of
terms start contributing at the same order. One should thus not be surprised to
find a  ground-state for $p=1$ which is not in the vortex-free sector \cite{Kells08}.

\subsection{Counting of states}

Before we turn to a detailed analysis of each QP subspace, let us show that we
do not miss any state using  simple counting arguments.
We have already seen in Sec.~\ref{sec:model}, that one has $N+2$ conserved
$\mathbb{Z}_2$ quantities (two loop operators, and $N$ plaquette operators),
with the constraint $\prod_p W_p=\mathbb{I}$. There is in fact one more
relation between the $W_p$'s, involving the number of bosons, which reads
%
%
\begin{equation}
    \label{eq:prodW_white}
    \prod_{p\in\mbox{white}}W_p=(-1)^{\sum_\bi\crb_\bi\anb_\bi}
    =(-1)^Q=\prod_{p\in\mbox{cyan (gray)}}W_p,
\end{equation}
%
%
showing that the parity of the number of vortices living on white plaquettes
(see Fig.~\ref{fig:pbc}) has to be the same as the parity of the number
of bosons. The last equality simply comes from the previously mentioned
constraint (\ref{eq:prodW}).
The first equality can be checked using the expression (\ref{eq:Wp_eff}) of the
$W_p$'s. Indeed, for a site $\bi$ having a white plaquette on its left, and
another one on its right, the product of the two associated $W_p$'s will give
$\tau_\bi^y\times(-1)^{\crb_\bi\anb_\bi}\tau_\bi^y=(-1)^{\crb_\bi\anb_\bi}$. In
the same way, for a site $\bi$ having a white plaquette above it, and another
one under it, the product of the two associated $W_p$'s will give
$\tau_\bi^z\times(-1)^{\crb_\bi\anb_\bi}\tau_\bi^z=(-1)^{\crb_\bi\anb_\bi}$.
Let us note that (\ref{eq:prodW_white}) has a meaning in the two bases we are
working in, the initial one and the unitarily transformed one. Indeed, in the
initial basis, the Hamiltonian $H$ commutes with the parity operator $(-1)^Q$~;
in the rotated basis, $H_\mathrm{eff}$ commutes with $Q$.

We thus see that in a subspace with a given number of QP's, there are $N$
independent conserved $\mathbb{Z}_2$ quantities. Thus, $N$ being the number of
effective spins $\tau_\bi$, there is no remaining effective spin degree of
freedom once the $\mathbb{Z}_2$ quantities are chosen. As a conclusion, the
$q$-QP subspace has dimension $d_q=2^N\begin{pmatrix}N\\q\end{pmatrix}$, with
the usual notation for binomial coefficients. This shows that we miss no state,
since
%
%
\begin{equation}
    \label{eq:counting}
    \sum_{q=0}^{N}d_q=2^N\sum_{q=0}^{N}\begin{pmatrix}N\\q\end{pmatrix}
    =2^{2N}=2^{N_\mathrm{s}},
\end{equation}
%
%
where $N_\mathrm{s}$ is the total number of spins in the brick-wall lattice.

This discussion furthermore sheds light on the fact that in order to compute
eigenenergies, a perturbative expansion of the Kitaev model (as opposed to exact
numerics) is really of interest only in the $0$-QP subspace. Indeed, we have
just seen that there are $N$ independent $\mathbb{Z}_2$ conserved quantities. It
is thus clear that as soon as we will have written down the effective Hamiltonian
in the $0$-QP subspace, the Hamiltonian will already be diagonal, whatever
the vortex configuration, although writing down the eigenstates of the
$\mathbb{Z}_2$ quantities in the basis of effective spin operators still has to
be done. However, in the $1$-QP subspace, one will have to diagonalize an
$N\times N$ matrix (numerically in the case of a nonperiodic vortex
configuration), which is identical to what one has to do when solving the
problem exactly as Kitaev did.
For $q\geqslant 2$, the perturbative expansion looks even more complicated than
the exact solution, but this is an artifact, since we recover free fermions.

%
%
\section{Effective Hamiltonian in the 0-QP subspace}
\label{sec:zeroqp}
%
%

\subsection{Effective Hamiltonian and eigenenergies}

In the 0-QP sector, and in the thermodynamical limit the effective Hamiltonian
(\ref{eq:heff_explicit}) simplifies and reads
%
%
\begin{equation}
    \label{eq:hamzeroqp}
    H_\mathrm{eff}|_{q=0} = E_0-\sum_{\{p_1,\ldots,p_n\}}
    C_{p_1,\ldots ,p_n} W_{p_1}\ldots W_{p_n},
\end{equation}
%
%
where $\{p_1,p_2,\ldots,p_n\}$ denotes a set of $n$ plaquettes and the
$W_p$'s are the conserved plaquette operators introduced in
Sec.~\ref{sec:model}. Note than when restricted to the 0-QP sector they simplify
to $\left.W_p\right|_{q=0}=\tau_{\mbox{\tiny L}}^y\,\tau_{\mbox{\tiny U}}^z\,
\tau_{\mbox{\tiny R}}^y\,\tau_{\mbox{\tiny D}}^z$
[see Eq.~(\ref{eq:Wp_eff})].

As mentioned at the end of the previous section, obtaining eigenenergies only
requires a minimal amount of work, namely replacing each $W_p$ by numbers
$w_p=\pm 1$, and doing the same with loop operators, without forgetting
about the constraints among these quantities.
The perturbative expansion of the
coefficients $E_0$ and $C_{p_1,\ldots ,p_n}$ are given in Appendix
\ref{app:zeroqp}.
Let us note that $\{p_1,p_2,\ldots,p_n\}$ does not need to be a linked cluster
of plaquettes (as seen for $C_{p,p+2n_1}$ that is nonvanishing at order
10), and that translational invariance of the Hamiltonian implies that the
$C_{p_1,\ldots ,p_n}$ coefficients only depend on $n-1$ relative positions of
the plaquettes.

%
%
\begin{figure}[t]
  \includegraphics[width=0.8\columnwidth]{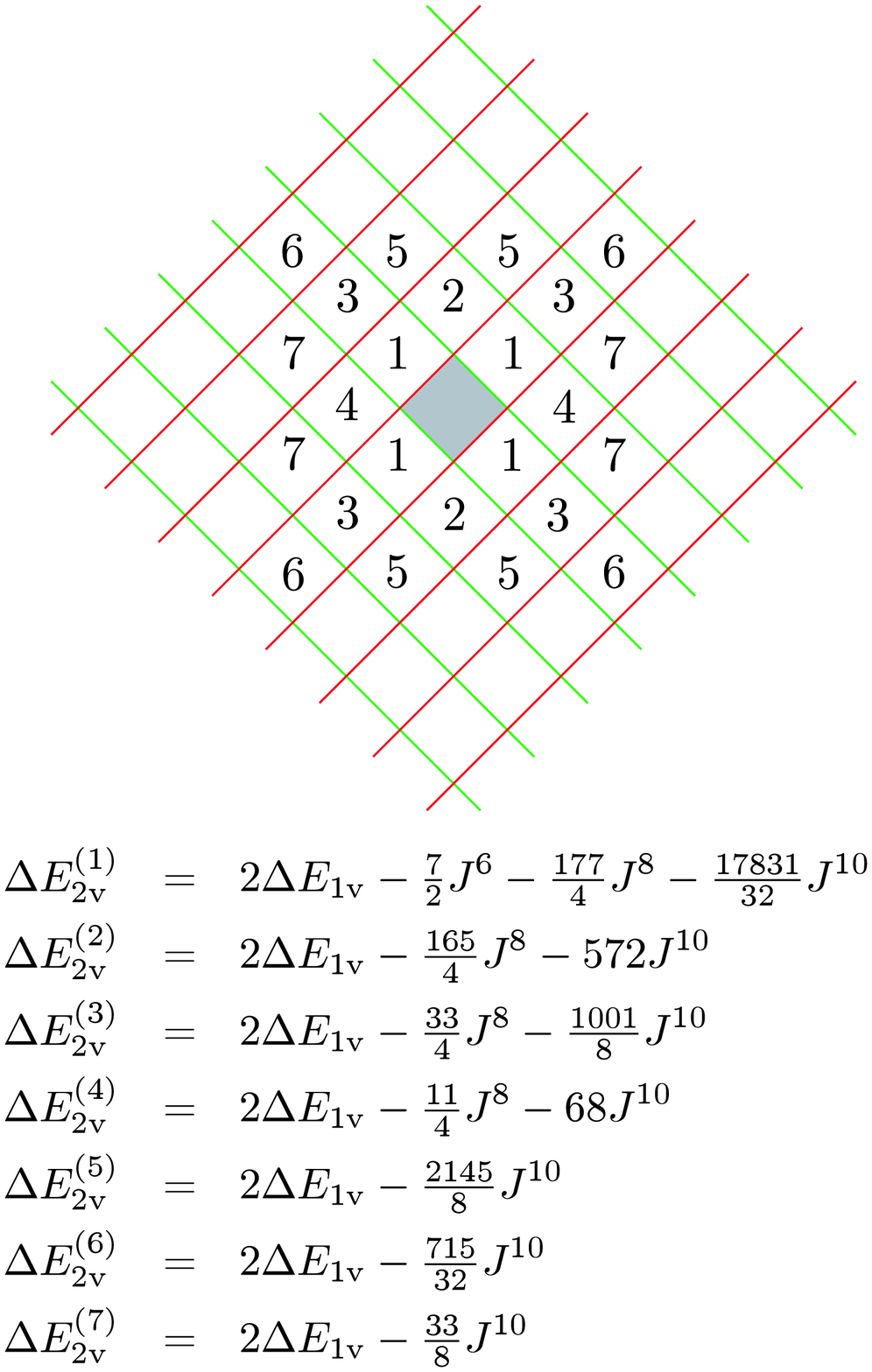}
  \caption{(color online). Two-anyon configurations (gray central plaquette
    and one of the numbered plaquettes) on a vortex-free background.
    $\Delta E_{1\mathrm{v}}$ ($\Delta E_{2\mathrm{v}}$) is the energy cost
    (at order 10) for adding one vortex (two vortices) to the vortex-free
    state.
    $\Delta E_{1\mathrm{v}}$ reads
    $\Delta E_{1\mathrm{v}}=J^4+8J^6+75J^8+784J^{10}$.
    For simplicity, we have set here $J_x=J_y=J$ but the results, in the
    general case, are easily obtained from the coefficients given in Appendix
    \ref{app:zeroqp}.}
  \label{fig:vv}
\end{figure}
%
%

The lowest nontrivial order involving the $W_p$'s (order 4) has been derived by Kitaev \cite{Kitaev06} ($C_p=J_x^2 J_y^2 /2$) and led him to identify the effective low-energy theory with the toric code \cite{Kitaev03}.
One of the main results of our work is to show that, at order 6 and beyond,
one obtains a {\em multi-plaquette expansion} in the effective low-energy
Hamiltonian. In other words vortices interact, though they remain static as
they have to since the $W_p$'s are conserved. The interaction energies between
vortices are not directly the $C$ coefficients. One should write $w_p=1-2n_p$
where $n_p$ is the number of vortices at plaquette $p$ ($0$ or $1$), then 
look at coefficients in the expansion in terms of the $n_p$'s.  The results of
such an analysis for two-vortex interaction energies in the case $J_x=J_y=J$ are
illustrated in Fig.~\ref{fig:vv} which shows that the interaction
{\em (i)} lowers the energy and is therefore attractive, {\em (ii)} is anisotropic even for $J_x=J_y=J$
which is clear from the structure of the underlying brick-wall lattice,
{\em (iii)} decreases with the distance $d$ between vortices as expected in a
gapped system. Note that for a finite-size system with PBC, the two-vortex
configurations with a central vortex and another vortex at one of 1, 5, 6 and
7 sites (see Fig.~\ref{fig:vv}) are forbidden since they violate the
constraint (\ref{eq:prodW_white}). A one-vortex configuration is also
forbidden since it violates the constraint (\ref{eq:prodW}). These
configurations would be allowed in an infinite system, or in a finite system
with open boundary conditions.

A most remarkable point which emerges from the analysis of $H_\mathrm{eff}|_0$
is that its eigenstates are those of the $W_p$'s. They are thus the same at any
order $(\geq 4)$, and are those of the toric code\cite{Kitaev03}, although their
eigenenergy changes with the perturbation order (we emphasize we are talking
about eigenstates of $H_\mathrm{eff}|_0$, not of the original Hamiltonian $H$). We graphically sketch the construction of these
eigenstates in the next subsection, which will also prove to be useful for the
$(q\geqslant 1)$ QP sectors, and show explicitly that they obey anyonic, more
precisely semionic, statistics. Our discussion of the toric code focuses on
peculiarities related to our way of studying the problem, that is not
restricted to the 0-QP subspace. For more details about the toric code model,
we refer the interested reader to
Refs.~\onlinecite{Kitaev03,Kitaev06,Preskill_HP}.

%
%
\subsection{The Toric Code in a nutshell}
\label{sec:sub:toricode}
%
%

\subsubsection{Mapping to the toric code}

%
%
\begin{figure}[t]
  \includegraphics[width=\columnwidth]
  {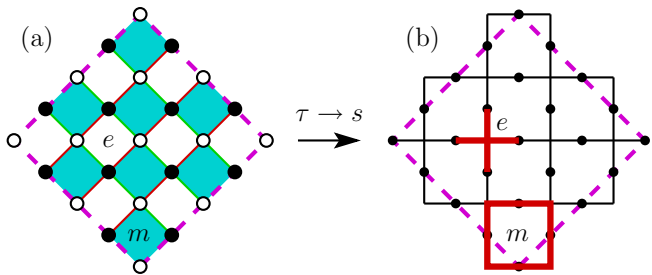}
    \caption{(color online). Illustration of the two different points of view
    one can have of a bi-colored lattice : sites are at vertices in (a) and on
    the bonds in (b). The plaquette $m$ in (a) remains a plaquette in (b), while
    the plaquette operator $e$ transforms into a star operator.}
    \label{fig:mapping_to_toric_code_lattice}
\end{figure}
%
%

As we have seen in the previous sections, the eigenstates of the effective
Hamiltonian in the 0-QP subspace are the eigenstates of the $W_p$'s and
of the $\mathcal{L}$'s. We recall that in this subspace, the plaquette
operators read $W_p|_{q=0}=\tau_{\mbox{\tiny L}}^y\,\tau_{\mbox{\tiny U}}^z\,
\tau_{\mbox{\tiny R}}^y\,\tau_{\mbox{\tiny D}}^z$ (see
Eq.~(\ref{eq:Wp_eff}) and Fig.~\ref{fig:mapping_brickwall_square}b).
A similar simplification occurs for the $\mathcal{L}$'s.
As mentioned by Kitaev\cite{Kitaev06} (for the Hamiltonian at order 4), the
effective Hamiltonian could be studied directly, but it is much easier to
visualize the eigenstates by performing some spin rotations, and bring the
Hamiltonian to the one of the toric code (generalized by multi-vortex terms).
Thanks to the special PBC we have chosen, the lattice sites can be bi-colored
in black and white as illustrated in
Fig.~\ref{fig:mapping_to_toric_code_lattice}. Then, one performs a different
rotation on the two kinds of sites
%
%
\begin{equation}
\label{eq:rotation}
\begin{array}{lclcl}
    \tau^x_\bullet=s^y &,& \tau^y_\bullet=s^z &,& \tau^z_\bullet=s^x,\\
    \tau^x_\circ=-s^y &,& \tau^y_\circ=s^x &,& \tau^z_\circ=s^z.
\end{array}
\end{equation}
%
%
This way, a cyan (gray) (resp. white) plaquette such as $m$ (resp. $e$) in
Fig.~\ref{fig:mapping_to_toric_code_lattice}a transforms into a
plaquette (star) term 
$B_m=(-1)^{\crb_{\mbox{\tiny L}}\anb_{\mbox{\tiny L}}
+\crb_{\mbox{\tiny D}}\anb_{\mbox{\tiny D}}}\,
s_{\mbox{\tiny L}}^z\,s_{\mbox{\tiny U}}^z\,
s_{\mbox{\tiny R}}^z\,s_{\mbox{\tiny D}}^z$
(resp. $A_e=(-1)^{\crb_{\mbox{\tiny L}}\anb_{\mbox{\tiny L}}
+\crb_{\mbox{\tiny D}}\anb_{\mbox{\tiny D}}}\,
s_{\mbox{\tiny L}}^x\,s_{\mbox{\tiny U}}^x\,
s_{\mbox{\tiny R}}^x\,s_{\mbox{\tiny D}}^x$), as shown with thick (red) lines in
Fig.~\ref{fig:mapping_to_toric_code_lattice}b. We have kept track of the
phases involving boson numbers because our construction will be needed for
($q\geqslant 1$)-subspaces, but it is clear that they can be dropped
{\em in the 0-QP subspace}.
Let us mention that the distinction between plaquette and star terms is purely
conventional.
The letters $m$ and $e$ refer to the magnetic and electric
vocabulary also used by Kitaev\cite{Kitaev06}, although we emphasize there is
absolutely no difference between an $A$ and a $B$ operator, which are both
disguised $W$ operators.
Up to an additive constant term, the effective Hamiltonian in the 0-QP subspace,
and at order 4 finally reads (with $J_\mathrm{eff}=J_x^2J_y^2/2$)
%
%
\begin{equation}
    \label{eq:Heff4}
    H_\mathrm{eff}|_{q=0}=-J_\mathrm{eff}\left(\sum_e A_e+\sum_m B_m\right).
\end{equation}
%
%
We work with this lowest (nontrivial) order Hamiltonian, because the
eigenstates of $H_\mathrm{eff}|_{q=0}$ remain the same whatever the
order in perturbation. One should simply remember that the eigenstates of $H$
also have to be eigenstates of $\mathcal{L}$ operators.
If the PBC are not of the type we use (see Fig.~\ref{fig:pbc}), the sites
can usually not be bi-colored and the rotations (\ref{eq:rotation}) cannot be
performed, which makes the construction much more complicated, and we shall
refer the interested readers to Ref.~\onlinecite{Wen03}.

%
%
\begin{figure}[t]
  \includegraphics[width=\columnwidth]{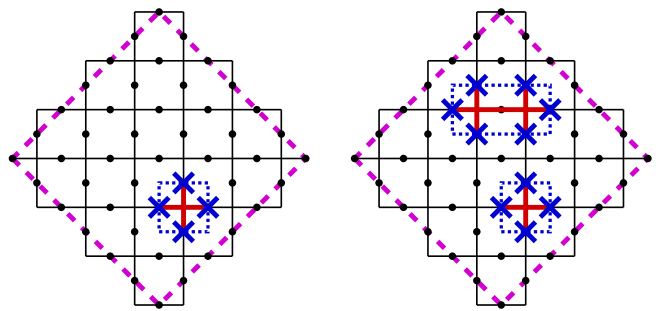}
    \caption{(color online). Two of the states entering the equal-weight
    superposition in (\ref{eq:gs}). Crosses on the sites indicate a spin
    flip, with respect to the reference state $|\!\Uparrow\rangle$ where all
    spins point upwards (the $A_e$ operators are still pictured by thick crosses at the vertices). The loops, in dotted lines, are obtained by
    joining the flipped spins.}
    \label{fig:Ae_loops}
\end{figure}
%
%

\subsubsection{Construction of the ground state(s)}

As a warming up, let us construct a ground-state of $H_\mathrm{eff}|_0$,
{\it i.~e.}, an eigenstate of all $B_m$ and $A_e$ operators with eigenvalue $1$
(there are actually four of these). An eigenstate of all $B_m$'s is for example
the "reference" state $|\!\Uparrow \rangle$ where all spins $s$ point in the $+z$-direction,
such that for all $\bi$ one has $s_\bi^z|\! \Uparrow \rangle =|\!\Uparrow\rangle$.
This state is not an eigenstate of the $A_e$ operators yet, but a simple
projection yields the desired state
%
%
\begin{equation}
    \label{eq:gs}
    2^{N/4-1/2}\prod_e\left(\frac{\mathbb{I}+A_e}{2}\right)
 | \!  \Uparrow \rangle\otimes|0\rangle_b,
\end{equation}
%
%
whose normalization follows from the number $N/2$ of $e$'s and the property
$\prod_e A_e=\mathbb{I}$ [see Eq.~(\ref{eq:prodW_white})].
The state $|0\rangle_b$ indicates that there is no quasiparticle, {\it i.~e.} no hardccore boson. A
graphical interpretation can be given of the state (\ref{eq:gs}) : it is an
equal-weight superposition of multi-loop configurations produced by the $A_e$
operators, as the ones shown in Fig.~\ref{fig:Ae_loops}.

\begin{figure}[t]
  \includegraphics[width=\columnwidth]{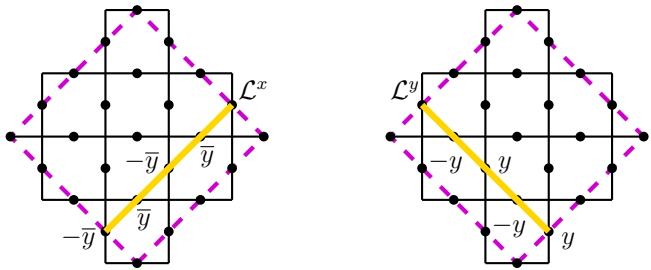}
    \caption{(color online). The two loop operators $\mathcal{L}^x$ and
    $\mathcal{L}^y$ and their contours in yellow (lightest gray) thick lines, corresponding
    to $\mathcal{C}_\mathrm{b}$ and $\mathcal{C}_\mathrm{c}$ in
    Fig.~\ref{fig:loop_operators_square}. As in the latter figure, but for $s$
    spins instead of $\tau$ spins, $y$ means $s^y$, etc. In the 0-QP sector,
    $\overline{y}=(-1)^{b^\dagger b}s^y$ is the same as $y$.}
    \label{fig:LxLy}
\end{figure}

One next has to get an eigenstate of two independent loop operators, which
we choose to be $\mathcal{L}_\mathrm{b}$ and $\mathcal{L}_\mathrm{c}$
(see Figs.~\ref{fig:loop_operators} and \ref{fig:loop_operators_square})
and which, from now on, will be denoted $\mathcal{L}^x$ and $\mathcal{L}^y$,
with eigenvalues $l^x$ and $l^y$.
The expressions of these operators in the $s$-spin language are given in
Fig.~\ref{fig:LxLy}.

\begin{figure}[t]
  \includegraphics[width=\columnwidth]{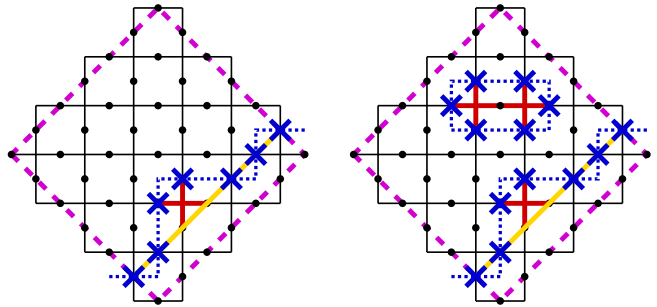}
    \caption{(color online). Two of the states entering the equal-weight
    (in absolute value) superposition in (\ref{eq:gs_lxly}), involving the
    $\mathcal{L}^x$ loop operator. Graphical conventions are the same as in
    Figs.~\ref{fig:Ae_loops} and \ref{fig:LxLy}.}
    \label{fig:Ae_Lx_loops}
\end{figure}

Note that in the 0-QP subspace, one could also have used other conserved loop
operators which are products of $s^x$ or of $s^z$ on the contours defining
$\mathcal{L}^x$ and $\mathcal{L}^y$. Such operators resemble more the ones
used by Kitaev\cite{Kitaev03}, but they are conserved only in the 0-QP
subspace (in contrast to $\mathcal{L}^x$ and $\mathcal{L}^y$), and so will not
prove to be very useful in the following.
%
%
%
%
As can be seen in Fig.~\ref{fig:LxLy}, $\mathcal{L}_x$ and $\mathcal{L}_y$
perform spin flips, with respect to $|\!\Uparrow\rangle$ on their associated
contours. The four ground-states of (\ref{eq:Heff4}) are then obtained with
another projection and proper normalization
%
%
\begin{eqnarray}
    \label{eq:gs_lxly}
    &&|\{w_p=1\},l^x,l^y \rangle_0=2^{N/4+1/2}\\
    && \times \left(\frac{\mathbb{I}+l^x\mathcal{L}^x}{2} \right)
    \left( \frac{\mathbb{I}+l^y\mathcal{L}^y}{2} \right)
    \prod_e\left(\frac{\mathbb{I}+A_e}{2}\right)
    |\! \Uparrow\rangle \otimes|0\rangle_b. \nonumber
\end{eqnarray}
%
%
These four states are equal-weight (in absolute value) superposition of all
possible multi-loop configurations, produced by the $A_e$ operators as in
Fig.~\ref{fig:Ae_loops}, as well as the $\mathcal{L}^x$ and $\mathcal{L}^y$
operators, as illustrated in Fig.~\ref{fig:Ae_Lx_loops} for $\mathcal{L}^x$.

%
%
%
%

Let us note that the preceding construction relies on the
$|\!\Uparrow\rangle$ state and the fact that it is an eigenstate of the
$B_m$'s, etc. However, one could also have started with a state
$|\!\Rightarrow\rangle$ where all spins point in the $x$-direction, which
is an eigenstate of the $A_e$'s and then follow a similar route.

\subsubsection{Construction of excited states}

We now have to see how to construct excited states, {\it i.~e.} states containing
vortices ($e$ or $m$) but still no quasiparticle. 

\begin{figure}[t]
  \includegraphics[width=\columnwidth]{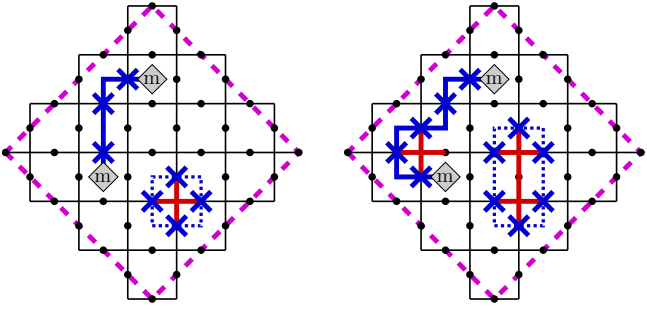}
    \caption{(color online). Two of the states entering the equal-weight
    (in absolute value) superposition for the state having two "magnetic"
    vortices. The latter are represented with little gray squares marked with
    the letter m, and are linked with a string of spin-flips [blue (dark gray) thick line].
    The other graphical conventions are the same as in
    Figs.~\ref{fig:Ae_loops}.}
    \label{fig:Ae_strings_loops}
\end{figure}

Constructing a state with some $B_m$'s being minus one ("magnetic vortices") is easy, once one has noticed that $s^x_\bi$ anticommutes with two $B_m$'s, and thus changes their values to their opposite.
Since $s^x_\bi$ commutes with all $A_e$'s, as well as with $\mathcal{L}^x$ and
$\mathcal{L}^y$ when $\bi$ does not belong to the corresponding contours,
$s^x_\bi|\{w_p=1\},l^x,l^y \rangle_0$ is an eigenstate of the effective Hamiltonian, with
two vortices living on the plaquettes touching the bond to which $\bi$ belongs. 
(Note that since $s^x_\bi$ also anticommutes with $\mathcal{L}^x$ and
$\mathcal{L}^y$ when $\bi$ belongs to the corresponding contours, one should use
a string of $s^x_\bj$ going around the torus without crossing $\mathcal{L}^x$
and $\mathcal{L}^y$ instead of $s^x_\bi$). The corresponding state is again an
equal-weight (in absolute value) superposition of states, but now with all
possible open strings joining the created vortices, as well as all possible
closed loops. This is illustrated in Fig.~\ref{fig:Ae_strings_loops}.

%
%
%
%

Creating "electric" vortices is as easy, since one can replace
$\prod_e\left(\frac{\mathbb{I}+A_e}{2}\right)$ in Eq.~(\ref{eq:gs_lxly}) by
$\prod_e\left(\frac{\mathbb{I}+a_eA_e}{2}\right)$, with $a_e=\pm 1$, respecting
the constraint $\prod_e a_e=1$ [see Eq.~(\ref{eq:prodW_white})]. Such a change can
also be obtained via the action of $s^z_\bi$ operators. Indeed, each $s^z_\bi$
operator anticommutes with two $A_e$'s, and thus changes their values to their
opposite. The fluctuation of the strings induced by $s^z_\bi$ operators is
however hard to see with the construction we have given, which relies on the
reference state $|\!\Uparrow\rangle$. To see this, one should construct
states from the reference state $|\!\Rightarrow\rangle$ where all spins point in
the $x$ direction, and then use projectors involving $B_m$'s instead of
$A_e$'s. This is not useful for our purpose so we let the interested reader
doing it on his own.

\subsubsection{The statistics of vortices}

\begin{figure}[t]
  \includegraphics[width=\columnwidth]{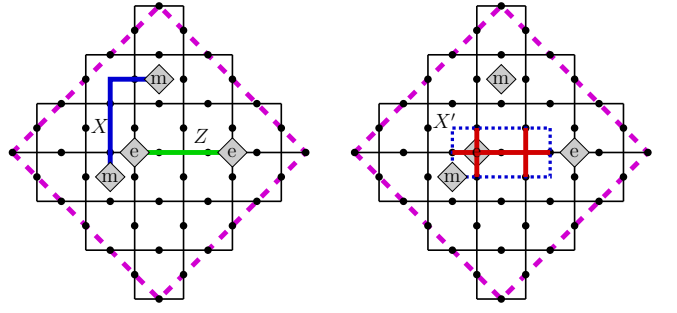}
    \caption{(color online). Illustration of operators involved in the braiding
    of a "magnetic" vortex around an electric vortex. $X$ is the product
    $\prod_\bi s^x_\bi$ for $\bi$ belonging to the blue (dark gray) thick path linking the two 
    m vortices (orthogonal to bonds). $Z$ is the product $\prod_\bi s^z_\bi$
    for $\bi$ belonging to the green (light gray) thick path linking the two e vortices (drawn on
    the bonds). $X'$ is the product $\prod_\bi s^x_\bi$ for $\bi$ belonging to
    the dashed loop around the leftmost e vortex, and is also equal to a
    product of the $A_e$'s operators encircled by the loop (denoted as thick
    crosses).}
    \label{fig:braiding_m_around_e}
\end{figure}

For completeness, let us now show that "magnetic" and "electric" vortices
behave as semions with respect to each other. This is done by first creating a
pair of "magnetic" vortices, then a pair of "electric" vortices, and finally
by moving one of the "magnetic" vortices around one of the "electric" vortices
as shown in Fig.~\ref{fig:braiding_m_around_e} (one could also do the contrary,
but then one should work with the reference state $|\!\Rightarrow\rangle$ to see
things more easily). With the notations of this figure (see also its caption), let us consider the
state $|\psi\rangle=ZX|\{w_p=1\},l^x,l^y \rangle_0$ with two e and m vortices.
Then the repeated application of spin-flips along the loop $X'$ (in any
direction) moves the downmost m vortex around the leftmost e vortex. The
resulting state is $X'|\psi\rangle$. But as $Z$ and $X'$ have one (and only one)
common site, they anticommute, whereas $X$ and $X'$ commute, so that
$X'|\psi\rangle=-ZXX'|\{w_p=1\},l^x,l^y \rangle_0$. Now, $X'$ which is a product of
$s^x_\bi$ operators forming a closed loop is nothing but a product of
$A_e$'s operators (the ones enclosed in the loop). As $|\{w_p=1\},l^x,l^y \rangle_0$
is an eigenstate of the $A_e$'s with eigenvalue one, we finally obtain that
\mbox{$X'|\psi\rangle=-|\psi\rangle$ :} braiding a magnetic vortex around an
electric vortex yields a nontrivial phase of $\pi$ ($-1=e^{\rmi\pi}$),
which proves the semionic statistics.

Let us mention that the magnetic vortices behave as bosons among themselves, and
so do the electric vortices. This is easily seen by noticing that creating and
moving $m$ vortices for example, only requires $s^x$ operators, which all
commute with one another.
To end this discussion about the statistics of vortices, let us also remark that
a compound object made of an electric and a magnetic vortex is a fermion (see
Ref.~\onlinecite{Preskill_HP}).

%
%
%
%

%
%
\section{Effective Hamiltonian in the 1-QP subspace}
\label{sec:oneqp}
%
%

\subsection{Form of the Hamiltonian}

The spectrum we obtained in the 0-QP subspace gives the lowest eigenenergies
for each configuration of the $W_p$'s. In this section, we explain how to
compute the high-energy spectrum for states with one quasiparticle, for each
$W_p$'s configuration, and how to build the associated eigenstates.
This is achieved by diagonalizing $H_\mathrm{eff}$ in the 1-QP subspace
(whose dimension is $d_1=N 2^N$, see the end of
Sec.~\ref{sec:perturbation}). In this subspace, the effective Hamiltonian
(\ref{eq:heff_explicit}) reads
%
%
\begin{eqnarray}
    \label{eq:ham1qp}
    H_\mathrm{eff}|_{q=1}&=&E_0+\mu-\sum_{\{p_1,\ldots,p_n\}}
    C_{p_1,\ldots ,p_n} W_{p_1}\ldots W_{p_n}\nonumber\\
    &&-\sum_{\{\bj_1,\dots,\bj_n\}} D_{\bj_1,\dots ,\bj_n}
    S_{\bj_1,\ldots,\bj_n}\crb_{\bj_n}\anb_{\bj_1},
\end{eqnarray}
%
%
where the second sum is performed over all non self-retracing paths of length $n$
starting at site $\bj_1$ and ending at site $\bj_n$, with possibly
$\bj_n=\bj_1$ when working at order 4 or higher. This is the reason why we
give the expansion up to this order but we would like to emphasize that obtaining orders up to 10 for
$H_\mathrm{eff}|_{q=1}$ is of the same complexity as for
$H_\mathrm{eff}|_{q=0}$. Self-retracing paths are renormalizing the chemical
potential $\mu$. Note that a hopping process of one quasiparticle
around a loop is nothing but the product of the $W_p$'s enclosed in the loop,
as can be easily checked. This explains why at order 4, one obtains some terms
proportional to $\crb_\bi \anb_\bi W_p$, where the plaquette $p$ shares site
$\bi$ (see Appendix \ref{app:oneqp}).

From now on ($q\geqslant 1$) the phase factors appearing in the $W_p$'s [see
Eq.~(\ref{eq:Wp_eff})] must be taken into account.
The operators $S$ have a structure similar to that of the $W_p$'s,
except that they are open string operators. They involve $\tau_\bj^\alpha$
as well as phase factors $(-1)^{\crb_\bj\anb_\bj}$ as follows
%
%
\begin{equation}
  \label{eq:Sn}
  S_{\bj_1,\ldots,\bj_n}=\varphi_{\bj_1,\ldots,\bj_n}
  {\mathcal T}_{\bj_{n-1}}^{\bj_n}\ldots {\mathcal T}_{\bj_1}^{\bj_2},
\end{equation}
%
%
where the $\varphi_{\bj_1,\ldots,\bj_n}$ are phase factors which reduce to the
identity in the $1$-QP subspace, and will be discussed later on
(see Sec.~\ref{sec:multiqps}). The two-site ${\mathcal T}_\bi^\bj$ operators are built
from the same $\tau_\bj^\alpha$ operators as the hoppings $t_\bi^\bj$, namely
%
%
\begin{eqnarray}
  \label{eq:S2}
  {\mathcal T}_\bi^{\bi+\bn_1}&=&\tau^x_{\bi+\bn_1}=({{\mathcal T}_{\bi+\bn_1}^\bi})^\dagger ,\\
  {\mathcal T}_\bi^{\bi+\bn_2}&=&-\rmi \tau^y_{\bi+\bn_2}\tau^z_\bi =({{\mathcal T}_{\bi+\bn_2}^\bi})^\dagger.
\end{eqnarray}
%
%
Note that in the $1$-QP subspace, one can also write the hopping term of the
Hamiltonian as\cite{Schmidt08}
%
%
\begin{equation}
  \label{eq:S_and_t}
  \left.S_{\bj_1,\ldots,\bj_n}\crb_{\bj_n}\anb_{\bj_1}
  \right|_{q=1}=t_{\bj_{n-1}}^{\bj_n}\ldots t_{\bj_1}^{\bj_2}.
\end{equation}
%
%

\subsection{Construction of a $1$-QP basis}

As can be seen when looking at the form of the hopping operators, bosonic and spin
degrees of freedom are coupled so that one has to tackle a polaron-like
problem. However, we shall now show that since all hopping operators
$t_\bi^\bj$ commute with all $W_p$'s as well as with all loop operators
$\mathcal{L}$, the $1$-QP problem is equivalent to that of one particle hopping
in a static magnetic field.

\begin{figure}[t]
  \includegraphics[width=\columnwidth]
  {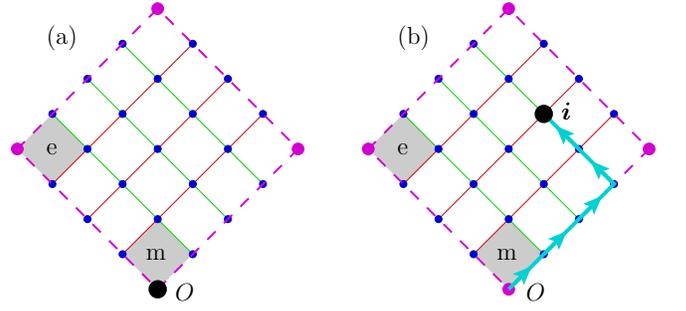}
    \caption{(color online). Illustration of states of the $1$-QP basis, built
      from the ground-state. The action of $\crb_O$ is to create a particle at
      site $O$ (large black filled circle) as well as to create a pair of e and
      m vortices (a). The action of the string operator $S$ represented with
      an oriented thick (cyan) line, on the state (a), then yields the state with
      the same vortices, but a particle at site $\bi$ (b). We do not bi-color the
      lattice in cyan (gray) and white any more, so that the figures are easier to
      read.}
    \label{fig:oneqp_basis}
\end{figure}

As a first step, we build a basis of the $1$-QP subspace. We denote by
$|\{w_p\},l^x,l^y\rangle_0$ a state of the $0$-QP subspace, which is an
eigenstate of the $W_p$'s and of $\mathcal{L}^x$ and $\mathcal{L}^y$, and built
as explained in Sec.~\ref{sec:sub:toricode}. We choose as the origin $O$ the
site we have already denoted with a large (magenta) filled circle 
(see Figs.~\ref{fig:pbc} and \ref{fig:loop_operators_square}, as well as
Fig.~\ref{fig:oneqp_basis}b). Let us then consider the state
$|\{w_p'\},l^x,l^y;O\rangle_1=\crb_O|\{w_p\},l^x,l^y\rangle_0$ belonging to
the $1$-QP subspace, and with one QP at the origin. From formula
(\ref{eq:Wp_eff}), it is clear that adding a particle at the origin changes the
value of two plaquettes, as illustrated in Fig.~\ref{fig:oneqp_basis}a for the
action of $\crb_O$ on the ground state, which is the reason why we made a
distinction between $w_p$ and $w_p'$. Note that all this is perfectly
consistent with Eq.~(\ref{eq:prodW_white}), as well as with the conclusion of
Ref.~\onlinecite{Levin03}. Indeed, in this paper, Levin and Wen showed that
fermions are always created in pairs, and this is the case here since a bound
object of an ``electric'' vortex and a ``magnetic'' vortex is a fermion (see
Sec.~\ref{sec:sub:toricode}) and our quasiparticles will turn out to be
fermions (see Sec.~\ref{sec:multiqps}).

%
%
%
%

Other states $|\{w_p'\},l^x,l^y;\bi\rangle_1$ with a particle at site $\bi$ are
obtained by applying an operator $S_{O,\ldots,\bi}\crb_\bi\anb_O$ onto $|\{w_p'\},l^x,l^y;O\rangle_1$ in order to make the particle hop, without affecting the conserved
$\mathbb{Z}_2$ quantities. 
Note that 
%
%
\begin{equation}
S_{O,\ldots,\bi}\crb_\bi\anb_O |\{w_p'\},l^x,l^y;O\rangle_1 =S_{O,\ldots,\bi}\crb_\bi |\{w_p\},l^x,l^y\rangle_0, 
\end{equation}
%
%

However, we still need a convention for the path to be taken
(which will amount to choose a gauge for the magnetic field the particles are
hopping in) to obtain a well-defined basis. 
The path from $O$ to $\bi$ is taken to first be in the $\bn_1$
direction as much as needed, then in the $\bn_2$ direction. For example, in
Fig.~\ref{fig:oneqp_basis}b, $\bi=3\bn_1+2\bn_2$ and the $S$ operator is
depicted as an oriented thick (cyan) line in this figure, with first 3 moves in
direction $\bn_1$ and then 2 moves in direction $\bn_2$.

\subsection{Hamiltonian in the $1$-QP basis}

\begin{figure}[t]
  \includegraphics[width=\columnwidth]
  {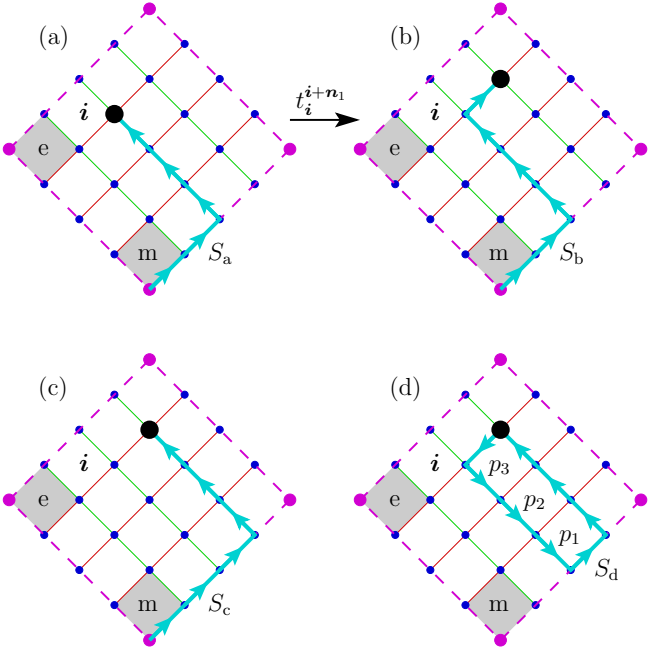}
    \caption{(color online). States (a), (b), (c) and contour (d) needed to
      compute the matrix element (\ref{eq:matrix_elem_t}), see text.}
    \label{fig:hopping_n1}
\end{figure}

Let us now consider the effective Hamiltonian at order $1$, for which the
hopping part is nothing but $T_0$, and study its action on a state
$|\{w_p\},l^x,l^y;\bi\rangle_1$. From the way the states have been built,
it is obvious that
$t_\bi^{\bi+\bn_2}|\{w_p\},l^x,l^y;\bi\rangle_1
=|\{w_p\},l^x,l^y;\bi+\bn_2\rangle_1$, and for the same reason and the fact
that $\mathcal{T}_{\boldsymbol{i}}^{\boldsymbol{j}}$ is unitary,
$t_\bi^{\bi-\bn_2}|\{w_p\},l^x,l^y;\bi\rangle_1
=|\{w_p\},l^x,l^y;\bi-\bn_2\rangle_1$.
We then turn to the hopping term $t_\bi^{\bi+\bn_1}$ and study its action
on the state $|\{w_p\},l^x,l^y;\bi\rangle_1$. In other words, we wish to compute the
matrix element
%
%
\begin{equation}
  \label{eq:matrix_elem_t}
  A_\bi^{\bi+\bn_1}=\phantom{\rangle}_1\langle\{w_p\},l^x,l^y;\bi+\bn_1|\,
  t_\bi^{\bi+\bn_1}\,|\{w_p\},l^x,l^y;\bi\rangle_1.
\end{equation}
%
%
All needed states are represented in Fig.~\ref{fig:hopping_n1} :
$|\{w_p\},l^x,l^y;\bi\rangle_1$ in (a),
$t_\bi^{\bi+\bn_1}\,|\{w_p\},l^x,l^y;\bi\rangle_1$ in (b) and
$|\{w_p\},l^x,l^y;\bi+\bn_1\rangle_1$ in (c). 
%
%
%
%
Using the notations of Fig.~\ref{fig:hopping_n1} ($S_\mathrm{a}$ is the oriented
product of spin operators $\mathcal{T}_\bi^\bj$ on the contour shown in (a), starting at
the origin, the same for $S_\mathrm{b}$ and $S_\mathrm{c}$, but for
$S_\mathrm{d}$ the product starts and ends at the particle's position),
it is easy to see that
$S_\mathrm{d}t_\bi^{\bi+\bn_1}\,|\{w_p\},l^x,l^y;\bi\rangle_1
=|\{w_p\},l^x,l^y;\bi+\bn_1\rangle_1$. Then, using the fact
that $S_\mathrm{d}^2=\mathbb{I}$,
%
%
\begin{equation}
  \label{eq:matrix_elem_t2}
  A_\bi^{\bi+\bn_1}
  =\phantom{\rangle}_1\langle\{w_p\},l^x,l^y;\bi+\bn_1|\,
  S_\mathrm{d}\,|\{w_p\},l^x,l^y;\bi+\bn_1\rangle_1,
\end{equation}
%
%
Furthermore a calculation on Pauli matrices shows that the action of
$S_\mathrm{d}$ on the state $|\{w_p\},l^x,l^y;\bi+\bn_1\rangle_1$ is the same
as the product of the plaquette operators encircled by the closed contour of
$S_\mathrm{d}$, which on the example of Fig.~\ref{fig:hopping_n1} reads
$W_{p_1}W_{p_2}W_{p_3}$. We finally obtain
%
%
\begin{equation}
  \label{eq:matrix_elem_t3}
  A_\bi^{\bi+\bn_1}=\prod_{p\subset\mathcal{E}_{\bi,\bi+\bn_1}} w_p,
\end{equation}
%
%
where the product has to be taken over all encircled plaquettes
$\mathcal{E}_{\bi,\bi+\bn_1}$ as illustrated on a particular example in
Fig.~\ref{fig:hopping_n1}.
The case of a hopping in the $-\bn_1$ direction can of course be deduced from
the above matrix elements by hermitian conjugation.

\begin{figure}[t]
  \includegraphics[width=\columnwidth]
  {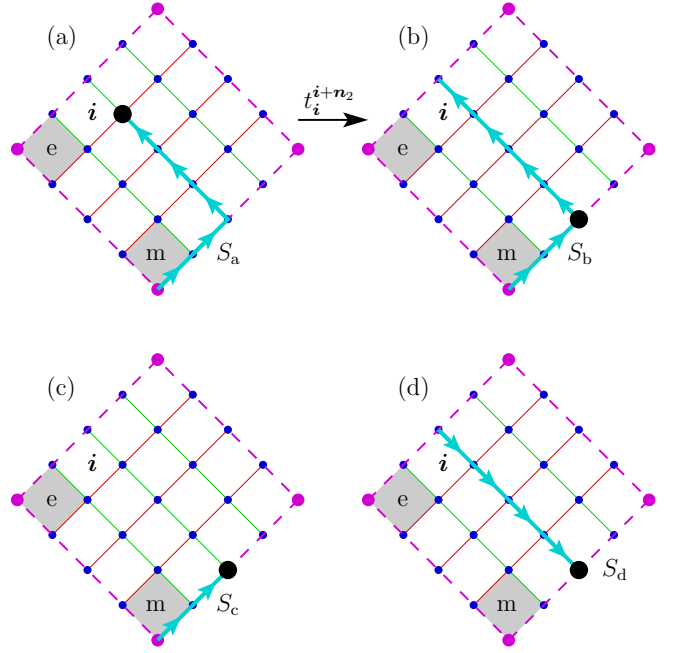}
    \caption{(color online). States (a), (b), (c) and contour (d) needed to
      compute the matrix element of a hopping in the $\bn_2$ direction,
      when site $\bi$ is ``one site away from the edge'' of the dashed (magenta)
      square, and thus involving a loop operator (d). See text for details.}
    \label{fig:hopping_n2_torus}
\end{figure}

For some hopping processes, the matrix element not only involves a product of
$w_p$'s but also a loop operator around the torus. This is illustrated in
Fig.~\ref{fig:hopping_n2_torus} for a hopping in the $\bn_2$ direction, starting
from site $\bi=i_x\bn_1+(2p-1)\bn_2$ ($i_x=2$ and $p=2$ in the figure).
%
%
%
%
In this case, one has $S_\mathrm{d}=\mathcal{L}^y_{i_x}$ ({\it i.~e.}, the loop operator in the $y$-direction, around the torus, going through the sites $\bi=i_x\bn_1+p\bn_2$ where $p$ takes all possible values), so that
%
%
\begin{equation}
  A_\bi^{\bi+\bn_2}=\phantom{\rangle}_1\langle\{w_p\},l^x,l^y;\bi|\,
  t_\bi^{\bi+\bn_2}\,|\{w_p\},l^x,l^y;\bi\rangle_1
  =l^y_{i_x}.
\end{equation}
%
%
As explained in Secs.~\ref{sec:model} and \ref{sec:mapping} the value of
$l^y_{i_x}$ is determined from the one of $l^y=l^y_1$ and from the value of the
plaquettes in between these two loop operators.

All the above examples lead to the following conclusion. The matrix elements of
the effective Hamiltonian at lowest order, in the $1$-QP subspace, are the same
as what one would obtain for a particle with hopping amplitudes $-J_x$ and
$-J_y$ in the $\bn_1$ and $\bn_2$ direction of the square lattice, in a
magnetic field whose (reduced) fluxes in plaquettes or cycles around the torus are
$\Phi/\Phi_0=0$ or $\Phi/\Phi_0=1/2$ (where $\Phi_0$ is the flux quantum).
This comes from the fact that, for example,
hopping around a plaquette $p$ gives a phase factor $1$ for the two hoppings in
the $\pm\bn_2$ directions, and an overall $w_p$ for the hoppings in the
$\pm\bn_1$ directions. The overall contribution is then $w_p$, which takes value
$w_p=\exp[2\rmi \pi \Phi/\Phi_0]$. This analysis can be extended to the case of hoppings
of the kind represented in Fig.~\ref{fig:hopping_n2_torus} where the PBC play a
role.

When tackling higher-order corrections, hoppings become longer-ranged as seen
in Eq.~(\ref{eq:ham1qp}), but the above considerations still apply because of 
Eq.~(\ref{eq:S_and_t}). It is then easy to compute the 1-QP spectrum for a
given map of the $\mathbb{Z}_2$ conserved quantities. 
As already explained, when the map does not possess translational invariance, one can only compute the spectrum numerically. When the $w_p$'s are translationally invariant, an
analytic solution is available, and for example, in the vortex-free subspace,
the dispersion relation obtained at order 2 (see Appendix \ref{app:oneqp}) is
%
%
\begin{eqnarray} 
  E^{\rm free}(k_x,k_y)&=&1-2[J_x \cos(k_x)+J_y \cos(k_y)] \\
  &&+2[J_x \sin(k_x) + J_y \sin(k_y) ]^2,\nonumber
\end{eqnarray} 
%
%
where the wave vector $(k_x,k_y)$ belongs to $[-\pi,\pi] \times [-\pi,\pi]$. The
gap in this sector is then obtained by minimizing $E_{\rm free}$ which yields
$\Delta^{\rm free}=1-2(J_x+J_y)$. Note that, in this case, the perturbative
result at order 1 coincides with the nonperturbative result obtained by Kitaev
\cite{Kitaev06} (see also Sec.~\ref{sec:checks}) and one recovers the transition
point at $J_x+J_y=1/2=J_z$.

Results for other sectors (vortex-full, or one vortex every two plaquettes) can
also be obtained. We mainly used them to check the validity of the coefficients
we computed perturbatively, as explained in Sec.~\ref{sec:checks}.

As a final remark about the $1$-QP subspace, let us mention a difference with what
is obtained when using exact fermionization methods. With these methods, the
low-energy subspace already contains many fermions, and one then considers
fermionic excitations on top of this complicated vacuum to reach high-energy
states. In our approach, the low-energy states are really empty of fermions, and
the excitations are only made of one particle, which can thus be qualified of
Landau quasiparticle.

%
%
\section{Effective Hamiltonian in the $(q\geqslant 2)$-QP subspace}
\label{sec:multiqps}
%
%

Let us now turn to multi-particle states with the aim of showing how the Fermi
statistics can be recovered from hardcore bosons with a string of spin and phase
operators. We shall not give many details here, since our approach becomes
cumbersome when studying multi-particle states, and because one knows from
exact solutions that one has to recover free fermions.

\subsection{Phase factors}

%
%
\begin{figure}[t]
  \includegraphics[width=0.7\columnwidth]
  {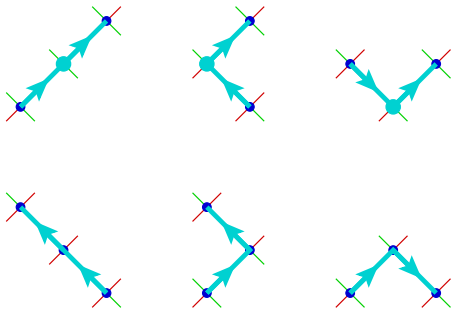}
  \caption{(color online). Illustration of the phase factors appearing at site $\bj_2$ in the
    string operators $S_{\ldots,\bj_1,\bj_2,\bj_3,\ldots}$.
    The phase factors $(-1)^{b^\dagger_{\bj_2} b_{\bj_2}}$
    are denoted as large (cyan) dots and are only involved in the three top
    processes which are, from left to right,
    $S_{\ldots,\bj,\bj+\bn_1,\bj+2\bn_1,\ldots}$,
    $S_{\ldots,\bj,\bj+\bn_2,\bj+\bn_1+\bn_2,\ldots}$ and
    $S_{\ldots,\bj,\bj-\bn_2,\bj+\bn_1-\bn_2,\ldots}$.}
  \label{fig:phase_factors}
\end{figure}
%
%

To obtain Fermi statistics, the phase factors appearing in the string operators
$S$ [see Eq.~(\ref{eq:heff_explicit})] are of utmost importance.
These factors do neither appear at the end sites of the string operators
$S$ [so in particular not at all for nearest-neighbor hoppings which are
simply the $t_\bi^\bj$ operators of Eq.~(\ref{eq:deftv})], nor at points of
turning back, but only at intermediate sites between two truly different other
sites. 
The six possibilities are shown in Fig.~\ref{fig:phase_factors}, where the phase factors occur only for the three topmost hoppings [phase factors for hoppings not represented in the figure can
be inferred from hermitian conjugation and Eq.~(\ref{eq:S2})]. In the figure,
 oriented thick (cyan) lines represent the string operators $S$, and the sites $\bj_2$
marked with a large dot are the ones involving a phase factor
$(-1)^{\crb_{\bj_2}\anb_{\bj_2}}$. As an example, the string operator associated
to a three-site hopping as the one shown top left in
Fig.~\ref{fig:phase_factors} reads
%
%
\begin{equation}
  \label{eq:ex_phase_factor}
  S_{\bj,\bj+\bn_1,\bj+2\bn_1}=(-1)^{\crb_{\bj+\bn_1}\anb_{\bj+\bn_1}}\,
  {\mathcal T}_{\bj+\bn_1}^{\bj+2\bn_1}\, {\mathcal T}_\bj^{\bj+\bn_1}.
\end{equation}
%
%

In fact, in $S_{\ldots,\bj_1,\bj_2,\bj_3,\ldots}$, a phase factor appears at
the intermediate site $\bj_2$ if ${\mathcal T}_{\bj_1}^{\bj_2}$ and ${\mathcal T}_{\bj_2}^{\bj_3}$
commute, and does not appear if they anticommute.

\subsection{Fermionic creation operators}

Rigorously, it is impossible to introduce creation/annihilation operators for
single fermions, because fermions should always be created/annihilated in
pairs. In fact, after choosing a site $O$ as an origin, and after choosing a
reference path from site $O$ to site $\bi$ (as was done in
Sec.~\ref{sec:oneqp}), the operator (running on this reference path)
$c^\dagger_\bi\sim S_{O,\ldots,\bi}^\mathrm{ref}\crb_\bi$ can be considered as a
fermionic creation operator at site $\bi$, once the origin $O$ has been sent to
infinity (using the same trick as when constructing a Dirac monopole in
electrodynamics). It should be clear from arguments similar to those of
Sec.~\ref{sec:oneqp} that such an operator creates a high-energy (spinless)
fermion at site $\bi$, but also creates (or destroys if there is already one)
one low-energy fermion made of two vortices, top and right of site $O$, as in
Fig.~\ref{fig:oneqp_basis}. It however commutes with all other $W_p$ operators,
except with these two.

The fermionic anticommutation relations between fermion operators at sites $\bi$
and $\bj$ can be checked by exhausting all possible crossings of two reference
paths $O,\ldots,\bi$ and $O,\ldots,\bj$. 

\subsection{Multi-particle basis and effective Hamiltonian}

From there on, one can construct a multi-particle basis of the Fock-space, as
was done for the one-particle basis, by successively creating fermions at some
sites (after having decided for an ordering of these sites).

It can then be shown, as was done in the 1-QP subspace, that the Hamiltonian is
nothing but a hopping Hamiltonian of free fermions in a magnetic field, whose
flux per plaquette is zero or half the flux quantum ($w_p=\pm 1$). The phase
factors, apart from ensuring proper Fermi statistics, also yield the correct
expressions for the $W_p$'s or product of $W_p$'s, which involve both $\tau$'s
and phase factors, and which appear for hoppings around closed paths.

\subsection{An alternative picture}

As was suggested by Levin and Wen in Ref.~\onlinecite{Levin03}, the statistics
of the effective quasiparticles can be probed with a simple argument. It
relies on exchanging two of these quasiparticles by using hoppings from the
Hamiltonian only, and doing so in such a way that a hopping on a bond between
two sites as occured exactly once in each direction, in order to capture phases
coming from the statistics only (and not, {\em e.~g.}, from a magnetic
Aharonov-Bohm-like phase).

\begin{figure}[t]
  \includegraphics[height=5cm]{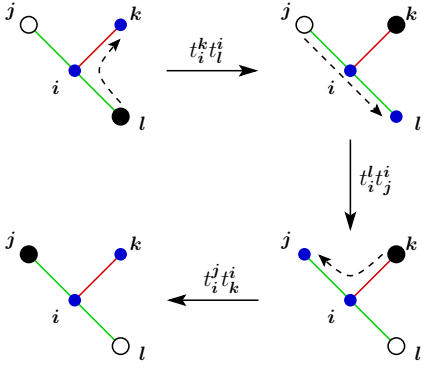}
  \caption{(color online). Illustration of the exchange of two particles discussed in the text,
    for $\bj=\bi+\bn_2$, $\bk=\bi+\bn_1$ and $\bl=\bi-\bn_2$.}
  \label{fig:exchange}
\end{figure}

Let us thus consider the exchange process of two particles initially sitting at
sites $\bj$ and $\bl$ (no other particle is present), as depicted in
Fig.~\ref{fig:exchange} and whose corresponding operator sequence is (using only
hopping operators arising at lowest order)
%
%
\begin{equation} 
  t^{\bj}_{\bi} t^{\bi}_{\bk} t^{\bl}_{\bi} t^{\bi}_{\bj} t^{\bk}_{\bi}
  t^{\bi}_{\bl}=-1,
\end{equation} 
%
%
or, equivalently,
$t^{\bi}_{\bj} t^{\bk}_{\bi} t^{\bi}_{\bl} =-t^{\bi}_{\bl} t^{\bk}_{\bi} t^{\bi}_{\bj}$. The sign in the latter identity confirms that
the quasiparticles made of a hardcore boson and an effective spin-1/2 obey
fermionic statistics.

%
%
%
%

%
%
\section{Simple checks from simple vortex configurations}
\label{sec:checks}
%
%

As shown by Kitaev \cite{Kitaev06}, the spectrum of the Hamiltonian
(\ref{eq:ham}) can be computed exactly by mapping the spin system onto free
Majorana fermions. The main drawback of this mapping is that one has first to
work in a fixed vortex-sector and, in a second step, perform the symmetrization
procedure involving all equivalent gauge sectors.  An alternative route
\cite{Feng07,Chen07_1,Chen08} consists in using the Jordan-Wigner
transformation which maps the problem onto free spinless fermions with $p$-wave
pairing. However, in both approaches and as is often the case, only periodic
configurations allow one to obtain analytical expressions of the spectrum.
In the following, we use Kitaev's approach (Majorana fermions) to compute the
spectrum in several simple periodic configurations characterized by a filling
factor $\nu=\frac{\rm Number\:\: of \:\: vortex}{\rm Number\:\: of \:\:
  plaquette}$.

Actually, diagonalizing the Majorana fermion Hamiltonian on this honeycomb
lattice \cite{Kitaev06} is completely equivalent to analyzing the problem of a
free particle on this lattice in a transverse magnetic field
\cite{Rammal85} with a flux per plaquette which can take only two values
corresponding to $w_p=\pm 1$ (see Appendix \ref{app:mapping} for details).
The ground-state energy is then obtained by filling all levels with negative
energy which amounts, in a bipartite lattice for which the spectrum is
symmetric, to consider half-filling.

For the three cases considered here, we compute the exact spectrum (still
assuming $J_z \geq J_x,J_y\geq 0$). Then, we perform the perturbative expansion
of the ground-state energy up to order 10.
This provides some simple checks of the results given in Sections
\ref{sec:zeroqp}.

%
%
\subsection{Vortex-free configuration $\nu=0$}
%
%
This configuration defined by $w_p=+1$ for all $p$'s is of special interest
since, in the thermodynamical limit, the ground state of $H$ lies in this
sector. This is a direct consequence of Lieb's theorem for flux phases
\cite{Lieb94}.
The spectrum, in this sector, is simply obtained since it is equivalent to
compute the spectrum of a free particle in zero field. The system being
periodic with $2$ sites per unit cell (see
Fig.~\ref{fig:mapping_brickwall_square}), the single-particle spectrum consists
of two bands given by the roots of the following characteristic polynomial
%
%
\begin{equation}
P^{\nu=0}(\varepsilon)=\varepsilon^2-f(\q)^2,
\end{equation}
%
%
where for all $\q$ in the reciprocal lattice, 
%
%
\begin{eqnarray}
  f(\q)^2&=&4\bigg\{J_x^2+J_y^2+J_z^2+2
  \bigg[J_x J_y \cos\Big({\q .(\n-\nn)}\Big)+\nonumber \\
  &&J_y J_z \cos(\q .\nn)+J_x J_z \cos(\q . \n)\bigg]  \bigg\}.
\end{eqnarray}
%
%
The ground-state energy per plaquette is thus given, in the thermodynamical
limit, by
%
%
\begin{equation}\label{eq:gsefree}
  e_0^{\nu=0}=-\frac{1}{8\pi^2} \int_{-\pi}^{\pi} {\rm d} q_x
  \int_{-\pi}^{\pi} {\rm d} q_y \:\: |f(\q)|.
\end{equation}
%
%
As already found by Kitaev, at the isotropic point \mbox{$J_x=J_y=J_z=1$},
one has $e_0^{\nu=0}\simeq-1.5746$.

The gap is given by the minimum, in modulus, of $P^{\nu=0}$'s roots,
{\it i.~e.}, $\min_\q |f(\q)|$. Thus, one obtains
%
%
\begin{equation}
  \label{eq:gapfree}
  \Delta^{\nu=0}=2(J_z-J_x-J_y).
\end{equation}
%
%

Setting $J_z=1/2$, and considering the perturbative limit $J_z\gg J_x,J_y$, one
obtains the following expansion for the ground-state energy at order 10
%
%
\begin{equation}
  e_0^{\nu=0}=-\frac{1}{2}-J^2-\frac{3 J^4}{4}-\frac{5 J^6}{2}
  -\frac{875 J^8}{64}-\frac{3087 J^{10}}{32}.
\end{equation}
%
%
For simplicity, we have set here $J_x=J_y=J$. This result can be easily
recovered by setting $w_p=+1$ for all $p$'s in Eq.~(\ref{eq:hamzeroqp}) and
using the coefficients given in Appendix \ref{app:zeroqp}.

One can also check directly the one-particle spectrum by expanding
$f(\q)$ in the same limit and by comparing it with the one-particle spectrum
in the vortex-free sector obtained in Sec.~\ref{sec:oneqp}.

%
%
\subsection{Vortex-full configuration $\nu=1$}
%
%

\begin{figure}[t]
  \includegraphics[width=0.8\columnwidth]{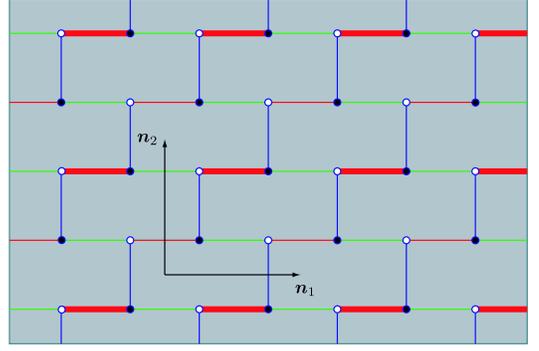}
  \caption{(color online). A possible gauge choice realizing the vortex-full
    lattice $\nu=1$. The thin (bold) links are associated to $u_{jk}=+1$
    ($u_{jk}=-1$) where $j$ belongs to the black sublattice and $k$ to the white
    one (see Appendix \ref{app:mapping}). The eigenvalue of the plaquette
    operator is then simply given by  $w_p=\prod_{(j,k) \in p}  u_{jk}$.}
  \label{fig:vortex1}
\end{figure}

The vortex-full sector is defined by $w_p=-1$ for all $p$'s.
In the ``particle in a field'' language, this problem corresponds to a magnetic
flux per plaquette which is half a flux quantum. With the gauge choice shown in
Fig.~\ref{fig:vortex1}, the system is periodic with $4$ sites per unit cell.
The single-particle spectrum thus consists of four bands given by the roots of
the characteristic polynomial
%
%
\begin{equation}
  P^{\nu=1}(\varepsilon)=\varepsilon^4-8 \, \varepsilon^2( J_x^2+J_y^2+J_z^2)
  + 16\,  g(\q)^2,
\end{equation}
%
%
where for all $\q$ in the reciprocal lattice, 
%
%
\begin{eqnarray}
  g(\q)^2&=&J_x^4+J_y^4+J_z^4-2\Big\{J_x^2 J_y^2 \cos(2 \q . \n )+\\
  &&J_y^2 J_z^2 \cos[\q . (\n-\nn)]-J_x^2 J_z^2 \cos[\q . (\n+\nn)]\Big\}.
  \nonumber 
\end{eqnarray}
%
%
The vectors $\bn_1=(1,0)$ and $\bn_2=(0,1)$ are defined in
Fig.~\ref{fig:vortex1}.
The ground-state energy per plaquette is  given, in the thermodynamical limit,
by
%
%
\begin{equation}\label{eq:gsefull}
  e_0^{\nu=1}=-\frac{\sqrt{2}}{8\pi^2} \int_{-\pi}^{\pi} {\rm d} q_x
  \int_{-\pi}^{\pi} {\rm d} q_y \sqrt{J_x^2+J_y^2+J_z^2 +|g(\q)|} ,
\end{equation}
%
%
Once again, for $J_x=J_y=J_z=1$, this expression gives
$e_0^{\nu=1}\simeq -1.5077$ in agreement with Kitaev's results \cite{Kitaev06}.

%
%
%
%

The gap is again given by the minimum, in modulus, of $P^{\nu=1}$'s roots
%
%
\begin{equation}
  \label{eq:gapfull}
  \Delta^{\nu=1}=2\bigg(J_z-\sqrt{J_x^2+J_y^2} \bigg),
\end{equation}
%
%
in agreement with results given in Ref.~\onlinecite{Pachos07}.

As for $\nu=0$, setting $J_z=1/2$, and considering the perturbative limit
$J_z\gg J_x,J_y$, one obtains the following expansion for the ground-state
energy at order 10
%
%
\begin{equation}
  e_0^{\nu=1}=-\frac{1}{2}-J^2+\frac{J^4}{4}-\frac{3 J^6}{2}+\frac{149 J^8}{64}
  -\frac{547 J^{10}}{32}.
\end{equation}
%
%

For simplicity, we have also set here $J_x=J_y=J$. This result can be easily
recovered by setting $w_p=-1$ for all $p$ in Eq.~(\ref{eq:hamzeroqp}) and using
the coefficients given in Appendix \ref{app:zeroqp}.

%
%
\subsection{Vortex-half configuration $\nu=1/2$}
\begin{figure}[t]
  \includegraphics[width=0.8\columnwidth]{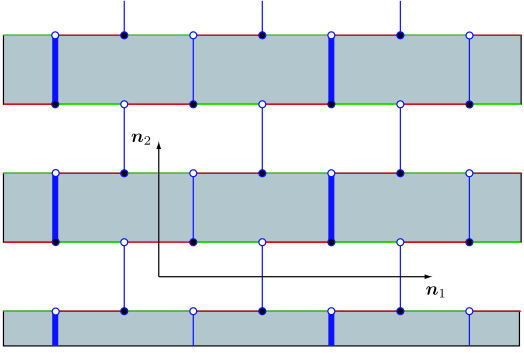}
  \caption{(color online). A possible gauge choice realizing the vortex-half
    lattice $\nu=1/2$. Notations are the same as in Fig.~\ref{fig:vortex1}. In
    this configuration, vortices are localized, in alternance on horizontal
    rows.}
  \label{fig:vortex2a}
\end{figure}

%
%
Let us now consider the vortex-half configuration shown in
Fig.~\ref{fig:vortex2a} which is  made of alternating vortex-free and
vortex-full rows. With the gauge choice shown in this figure, the system is
periodic with 8 sites per unit cell. The 8 bands of the single-particle
spectrum are given from the roots of the following characteristic polynomial
%
%
\begin{widetext}
  \begin{eqnarray}
    P^{\nu=1/2}(\varepsilon)&=&\varepsilon^8
    -16 \, \varepsilon^6 (J_x^2+J_y^2+J_z^2)+ 
    32 \,\varepsilon^4 \big[
    3(J_x^4+J_y^4+J_z^4) + 
    4( J_x^2 J_y^2+J_y^2 J_z^2+J_x^2J_z^2) -
    2J_x^2 J_y^2 \cos(\q . \n)
    \big] - \nonumber \\
    &&256 \,\varepsilon^2 \big[
    J_x^6+J_y^6+J_z^6+
    (J_x^2+ J_y^2)(J_y^2+ J_z^2)(J_x^2+ J_z^2)  - 
    2 J_x^2 J_y^2 (J_x^2+ J_y^2)  \cos(\q . \n) 
    \big]+\nonumber \\
    && 256\bigg\{J_x^8+J_y^8+J_z^8 +
    4 J_x^4 J_y^4-4(J_x^2 J_y^6+J_x^6 J_y^2) \cos(\q . \n)
    +2J_x^2\big[J_x^2 J_y^4\cos(2\q . \n)-J_y^2 J_z^4\cos(2\q . \nn)\big]
    +  \nonumber\\
    && 2J_z^4\big\{J_x^4 \cos[\q .(\n+2 \nn)]
    +J_y^4 \cos[\q .(\n-2 \nn)] \big\} \bigg\}\nonumber.
  \end{eqnarray}
\end{widetext}
%
%
The vectors $\bn_1=(1,0)$ and $\bn_2=(0,1)$ are defined in
Fig.~\ref{fig:vortex2a}.
Note that since the hexagonal lattice is bipartite, the single-particle
spectrum is even and, consequently,  all characteristic polynomials are
functions of $\varepsilon^2$. Thus, even in this vortex-half configuration, one
can get analytical expressions for the 8 bands since, practically, one only
has to find the roots of a fourth-order polynomial.

%
%
%
%

At the isotropic point, one obtains the ground-state energy per plaquette
$e_0^{\nu=1/2}\simeq -1.5227$. The gap is given by the minimum, in modulus, of $P^{\nu=1/2}$'s roots
%
%
\begin{equation}
  \label{eq:gapcheck}
  \Delta^{\nu=1/2}=2\bigg(J_z-\sqrt{J_x^2+J_y^2} \bigg).
\end{equation}
%
%
It is worth noting that the gap, in this sector is exactly the same as the one
in the vortex-full sector $\Delta^{\nu=1}$ [see Eq.~(\ref{eq:gapfull})].

Expanding the negative roots of $P^{\nu=1/2}$ at order 10 and integrating them
out as in the previous sector, one gets for $J_x=J_y=J$
%
%
\begin{equation}
  \label{eq:gsedemia}
  e_0^{\nu=1/2}=-\frac{1}{2}-J^2-\frac{J^4}{4}+\frac{3 J^6}{2}
  -\frac{411 J^8}{64}-\frac{211 J^{10}}{32}.
\end{equation}
%
%

\begin{figure}[t]
  \includegraphics[width=0.8\columnwidth]{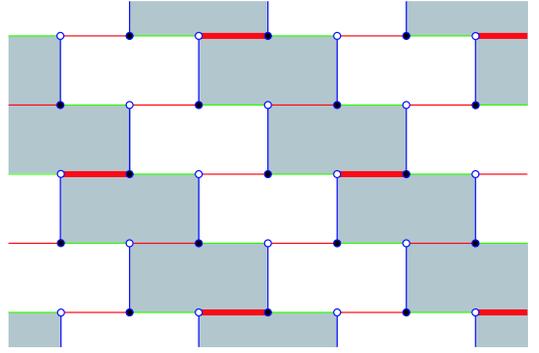}
  \caption{(color online). Another possible gauge choice realizing the
    vortex-half lattice $\nu=1/2$. Notations are the same as in
    Fig.~\ref{fig:vortex1}. In this configuration, vortices are localized, in
    alternance, on diagonal bands.}
  \label{fig:vortex2b}
\end{figure}

Finally, one may also consider another vortex-half configuration rotated as
shown in Fig.~\ref{fig:vortex2b}.
%
%
%
%
The corresponding characteristic polynomial is straightforwardly obtained from
$P^{\nu=1/2}$ by the permutation $J_x \rightarrow J_y$,  $J_y \rightarrow J_z$,
$J_z \rightarrow J_x$. However, since the perturbation is performed in the limit
$J_z \gg J_x,J_y$ it leads to a different expression for the expanded
ground-state energy. In this case, one gets for $J_x=J_y=J$
%
%
\begin{equation}
  \label{eq:gsedemib}
  e_0^{\nu=1/2}=-\frac{1}{2}-J^2-\frac{J^4}{4}-\frac{J^6}{4}+\frac{109 J^8}{64}
  +\frac{59 J^{10}}{16}.
\end{equation}
%
%

Once again, both expressions (\ref{eq:gsedemia}) and (\ref{eq:gsedemib}) can be
recovered from Eq.~(\ref{eq:hamzeroqp}) using the coefficients given in
Appendix \ref{app:zeroqp}.

The various results obtained for $\nu=0,1,1/2$ provide (partial) checks of the coefficients given in Appendices \ref{app:zeroqp} and \ref{app:oneqp} and show the power of the PCUTs to compute high-order expansion for the spectrum. In the next section, we shall show that this method is also an efficient tool to tackle more complex problematics.

%
%
\section{Observables}
\label{sec:Observables}
%
%

One of the advantages of the CUTs method is that it allows one to obtain the
effective form of any observables and to compute its matrix elements in the
eigenbasis of the Hamiltonian. 
The aim of this section is two-fold. First, we compute perturbatively the spin-spin correlations and show that they admit a plaquette-operator expansion similar to that of the spectrum. The second part of this section is dedicated to the most fundamental problem of local spin operations onto the ground state. Following Ref.~\onlinecite{Dusuel08_1}, we show that single-spin operations create anyons but also fermions. We compute the spectral weights of various states stemming from such operations and we also analyze the action of string operations which allow for manipulation of anyons. Finally, we give a procedure to derive the operators which create anyons without fermions and show that they involve tricky superpositions of multi-spin operators.

%
%
\subsection{Spin-spin correlation functions}
\label{sec:CF}
%
%

The Hamiltonian (\ref{eq:ham}) is invariant under the time-reversal symmetry since it is a quadratic function of the spin operators. Thus, any expectation value of an odd number of spin operators vanishes (such as the magnetization $\langle\sigma_\bi^\alpha\rangle$). Note also that the absence of odd cycles ensures that the eigenstates do not break this symmetry \cite{Kitaev06,Yao07,Dusuel08_2}.

In addition, the only nonvanishing spin correlators are those involving products of $\sigma^\alpha_\bi \sigma^\alpha_\bj$ on $\alpha$-dimers \cite{Baskaran07,Chen08,Yang08}.
In this section, we focus on the spin-spin correlation functions and their expression in the 0-QP sector. 
More precisely, we consider the following operators $C_{\bi,\bj}^{\alpha \alpha}=\sigma^\alpha_\bi \sigma^\alpha_\bj$ where $(\bi,\bj)$ is an $\alpha$-dimer. 
To compute these quantities, we proceed in a way similar to what we
have already done to derive the effective Hamiltonian : 
\begin{itemize}

\item  we express the observable in the ESB language~; 

\item  we compute its effective form perturbatively as explained in  Appendix \ref{app:observable} (see also Ref.~\onlinecite{Knetter04} for a detailed discussion)~; 

\item  we project it out in the sector of interest.

\end{itemize}

Using the ESB form of the spin operators (\ref{eq:mapping}), we
straightforwardly achieve the first step  mentioned above for the three
correlation functions
%
%
\begin{eqnarray} 
\sigma_{\bi,\circ}^x  \sigma_{\bi+\bn_1,\bullet}^x &=&
(\crb_\bi+\anb_\bi)\,\tau_{\bi+\bn_1}^x(\crb_{\bi+\bn_1}+\anb_{\bi+\bn_1}) ,\\
\sigma_{\bi,\circ}^y \sigma_{\bi+\bn_2,\bullet}^y&=&\mathrm{i} \, \tau^z_{\bi}
(\crb_\bi-\anb_\bi)\, \tau_{\bi+\bn_2}^y (\crb_{\bi+\bn_2}+\anb_{\bi+\bn_2}),\quad \\
\sigma_{\bi,\circ}^z \sigma_{\bi,\bullet}^z&=&
1-2 \crb_\bi\anb_\bi=(-1)^{\crb_\bi \anb_\bi}.
\end{eqnarray}
%
%
To avoid any ambiguity, we keep track of the type of sites ($\bullet$ or
$\circ$) but we are working, at this stage, on the effective square lattice.
Next, we turn to the second step using the perturbative expansion described in
Appendix \ref{app:observable}. In the present case, we pushed
the calculation up to order 6 and, finally, we focus on the 0-QP sector.

As one expects, the effective form of the spin-spin correlation function is
similar to that of the effective Hamiltonian. This is due to the fact that, in
the low-energy sector, $W_p$'s are the only degrees of freedom.
Thus, we obtain, an expansion in terms of the plaquette operators
%
%
\begin{equation}
  \label{eq:obs_expansion_0qp}
  C_{\bi \bj}^{\alpha \alpha}|_{q=0} = a^{\alpha \alpha}
  -\sum_{\{p_1,\ldots,p_n\}} b^{\alpha \alpha}_{p_1,\ldots ,p_n} W_{p_1}
  W_{p_2}\ldots W_{p_n}. 
\end{equation}
%
%
The coefficients $a^{\alpha \alpha}$ and $b^{\alpha \alpha}_{p_1,\ldots ,p_n}$
are given in Appendix \ref{app:CF} up to order 6. Here again, we can see that
these correlation functions involve interactions between connected or
disconnected plaquettes.

As a simple check of our expression, one can easily compute the nonperturbative
correlation function in the vortex-free and the vortex-full sector thanks to
the Hellman-Feynman theorem.
Indeed, in these sectors, all sites are equivalent so that one readily gets the expectation value
%
%
\begin{equation} 
  \langle C_{\bi \bj}^{\alpha \alpha}|_{q=0} \rangle_\nu=
 C_{\bi \bj}^{\alpha \alpha}|_{q=0}^\nu=-\frac{\partial e_0^\nu}{\partial J_\alpha}
\end{equation}
%
%
for both cases $\nu=0,1$ for which the ground-state energies are given in Eqs.~(\ref{eq:gsefree}) and (\ref{eq:gsefull}). As in the previous section, the subscript $\nu$ indicates that we consider the ground state of the sector with filling factor $\nu=0,1$.
Then, expanding these expressions (before derivation), setting $J_z=1/2$ and for simplicity $J_x=J_y=J$, one gets
%
%
\begin{eqnarray} 
  C_{\bi }^{zz}|_{q=0}^{\nu=0}&=&1-2 J^2-\frac{9 J^4}{2}-25 J^6, \\
  C_{\bi \bj}^{xx}|_{q=0}^{\nu=0}&=&
  J+\frac{3 J^3}{2}+\frac{15 J^5}{2}=C_{\bi \bj}^{yy}|_{q=0}^{\nu=0},
\end{eqnarray}
%
%
for the vortex-free sector and  
%
%
\begin{eqnarray} 
  C_{\bi }^{zz}|_{q=0}^{\nu=1}&=&1-2 J^2+\frac{3 J^4}{2}-15 J^6, \\
  C_{\bi \bj}^{xx}|_{q=0}^{\nu=1}&=&
  J-\frac{ J^3}{2}+\frac{9 J^5}{2}=C_{\bi \bj}^{yy}|_{q=0}^{\nu=1},
\end{eqnarray}
%
%
for the vortex-full sector. As can be checked, these results can be recovered
using the coefficients given in Appendix \ref{app:CF} and Eq.~(\ref{eq:obs_expansion_0qp}).
We emphasize that, as for the spectrum, our expressions allow us to investigate
arbitrary vortex configurations such as sparse vortex ones, recently studied
numerically \cite{Lahtinen08,Yu07}.

%
%
\subsection{Creation of anyons}
\label{subsec:creation_anyons}
%
%
Let us now analyze the action of a single-spin operation onto the ground state and following Ref.~\onlinecite{Dusuel08_1}, let us focus on $\sigma_{\bi,\bullet}^z$. As for the correlation functions, one first has to write this operator in the ESB formalism which is, again, straighforward since $\sigma_{\bi,\bullet}^z=\tau_{\bi}^z$. Then, one computes its renormalization under the unitary transformation $U$ which ``diagonalizes'' the Hamiltonian. Finally, one can compute any matrix element of this observable between any eigenstates.

\begin{figure}[t]
  \includegraphics[width=5cm]{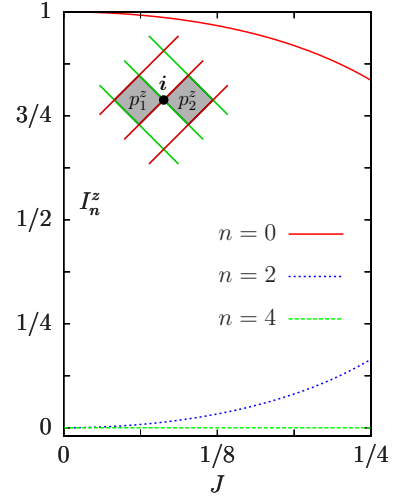}
  \caption{(color online).
    Behavior of the spectral weights $I_n^z$ 
    for fermion numbers $n=0,2,4$, as a function of the coupling $J=J_x=J_y$ for
    $J_z=1/2$. Gray plaquettes in the insets show the positions $p_1^{z}$ and
    $p_2^{z}$ at which the anyons are created under the action of
    $\tau_\bi^z$.}
  \label{fig:Iyz}
\end{figure}

At order 0, the observable is not renormalized and one has $U^\dagger \tau_\bi^z  U= \tau_\bi^z$. When this operator acts onto the ground state which is in the vortex-free sector, it thus simply flips the two plaquettes as shown in Fig.~\ref{fig:Iyz}. In other words, it creates two anyons and nothing else.

%
%
%
%

At order 1, one gets
%
%
\begin{equation}
  \label{eq:rensz1}
  U^\dagger \tau_\bi^z  U = \tau_\bi^z \left[1+
    \left(J_x v_{\bi-\bn_1}^{\bi} +J_y v_{\bi-\bn_2}^{\bi}
      +\mbox{ h.c.}\right)\right],
\end{equation}
%
%
showing that things are more complex since pairs of particles (fermions) are created. It means that, at this order, $\tau_\bi^z$ couples the 0-QP subspace of the vortex-free sector with the 2-QP subspace of the two-vortex sector discussed above.
To have a physical quantitative picture of such processes, let us compute  the spectral weights defined as
%
%
\begin{equation}
  \label{eq:weights}
  I^z_n =\sum_{\bk} \left|\langle \{p_1^z,p_2^z\},n, \bk | \tau_\bi^z |0\rangle\right|^2,
\end{equation}
%
%
where $|\{p\},n,\bk \rangle$ denotes the eigenstate of $H$ in a sector given by an anyon configuration $w_p=-1$, and $n$ high-energy quasiparticles with quantum numbers $\bk$. Here, the plaquettes 
$p_1^z$ and $p_2^z$ are as indicated in the inset of Fig.~\ref{fig:Iyz}.
This quantity measures the weight of all $n$-fermion contributions obtained by the action of $\tau_\bi^z$ onto the ground state $|0\rangle$ which contains no fermion and no anyon. As it should, these spectral weights satisfy the sum rule $\sum_n I^z_n=1$.
At order 6, one gets
%
%
\begin{eqnarray}
 I^z_0&=&1-\left(J_x^2+J_y^2 \right)-\frac{3}{2}\left(J_x^4+J_y^4\right)-4 J_x^2 J_y^2- \nonumber \\
 && \frac{7}{2} \left(J_x^6+J_y^6 \right)-\frac{43}{2} \left(J_x^2 J_y^4+J_x^4 J_y^2 \right),\\
 I^z_2&=&J_x^2+J_y^2 +\frac{3}{2}\left(J_x^4+J_y^4\right)+ 4 J_x^2 J_y^2+\nonumber \\
 && \frac{7}{2} \left(J_x^6+J_y^6 \right)+ \frac{43}{2} \left(J_x^2 J_y^4+J_x^4 J_y^2 \right),
\end{eqnarray}
%
%
which shows the importance of the two-fermion states for increasing couplings as can be seen in Fig.~\ref{fig:Iyz}.  Note that the sum rule is fulfilled here implying $I_{n\geq 4}^z=0$ at order 6.  Actually, one may consider representative curves in Fig.~\ref{fig:Iyz} as almost converged since order 8 corrections would bring very small corrections. 

To summarize, one must realize that local spin operations onto the ground state create anyons (here two) but also give rise to fermionic excitations whose weight increases significantly with the perturbation.

%
%
\subsection{Manipulation of anyons}
\label{subsec:manipulation_anyons}
%
%
Another important question concerns the manipulation of the anyons which, as
shown above, may be created by local spin operations. Such an issue is of
special interest for experiments aiming at braiding anyons \cite{Jiang08}.
This topic has been the subject of a recent controverse with  Zhang {\it et al.} \cite{Zhang08_1,Vidal08_1} who completely neglected the existence of fermions in this model.
Following Jiang {\it et al.} \cite{Jiang08} who proposed an ingenious protocol to detect anyons statistics, we wish to compute the action of a string operator onto the ground state. 

\begin{figure}[t]
  \includegraphics[width=\columnwidth]{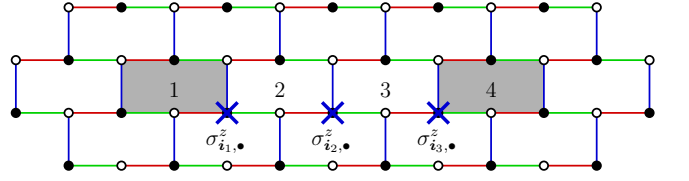}
  \caption{(color online).
Action, in the vortex-free sector, of the string operator $S=\sigma_{\bi_3,\bullet}^z \sigma_{\bi_2,\bullet}^z \sigma_{\bi_1,\bullet}^z $. Each operator flips the two plaquettes adjacent to the $z$-dimer it is attached to but also creates fermionic excitations (not shown).}
  \label{fig:braid_z}
\end{figure}

For simplicity, we consider here the operator \mbox{$S=\prod_{a=1,m} \sigma_{\bi_a,\bullet}^z$} along a horizontal line of the original brick-wall lattice (see Fig.~\ref{fig:braid_z} with $m=3$ for notations). 

At order 0, it is simple to see that $S$ first creates two anyons and make one of them jump in the direction of the string so that, at the end, one eventually has one anyon at plaquette 1, another anyon at the plaquette $m+1$, and no fermion. 
%
%
%
%

However, at higher orders, as previously, such an operation creates fermions. To quantify this phenomenon, we consider the probability ${\mathcal P}=\left|\langle \{1,m+1\},0 | S |0\rangle\right|^2$ to find the final state in the lowest-energy state (no fermion) with anyons at plaquettes 1 and $(m+1)$ which coincides with $I^z_0$ for $m=1$. 
In Ref.~\onlinecite{Dusuel08_1}, we computed this probability at order 2, but here, we go beyond and give the result at order 6 
%
%
\begin{eqnarray}
  \label{eq:proba}
{\mathcal P}&=&1-m\big(J_x^2+J_y^2 \big)+ \\
&& \frac{m(m-4)}{2} \big(J_x^4+J_y^4 \big)+  (m^2-8m+3) J_x^2 J_y^2- \nonumber \\
&& \frac{m(m^2-12m+32)}{6} \big(J_x^6+J_y^6 \big) +\nonumber \\
&& \bigg(-\frac{m^3}{2} +10 m^2-51 m+ 20\bigg) \big(J_x^4 J_y^2+J_x^2 J_y^4 \big) \nonumber.
  \end{eqnarray}
%
%

The main reason to perform this high-order calculation is that the above expression pleads in favor of an exponentiated form linear with $m$. Indeed, although we have no proof, we conjecture that ${\mathcal P}$ can be recast into $\exp(A- m B)$ as suggested in footnote 4 in Ref.~\cite{Kitaev06}. It is indeed striking to see that the expression (\ref{eq:proba}) which is a polynomial of the variable $m$, can be seen as the expansion of such a simple form with, at order 6
%
%
\begin{eqnarray}
A&=&3 J_x^2 J_y^2 + 20 (J_x^4 J_y^2 + J_x^2 J_y^4) ,\\
B&=& J_x^2 + J_y^2 + 8 J_x^2 J_y^2 + 2 (J_x^4+J_y^4) + \nonumber\\
&& 48 (J_x^4 J_y^2 + J_x^2 J_y^4)+\frac{16}{3} (J_x^6+J_y^6).
\end{eqnarray}
%
%
Further, it is clear that ${\mathcal P}$ is bounded by $0$ and $1$ for any $m$,
which is clearly not the case if one considers (\ref{eq:proba}). Let us also
note that the fact that $U^\dag \tau_\bi^z U$ is found to be proportional to
$\tau_\bi^z$ [see Eq.~(\ref{eq:rensz1})] strengthen the idea of an exponential
form of this effective observable and hence for $S$.

\begin{figure}[t]
  \includegraphics[width=7cm]{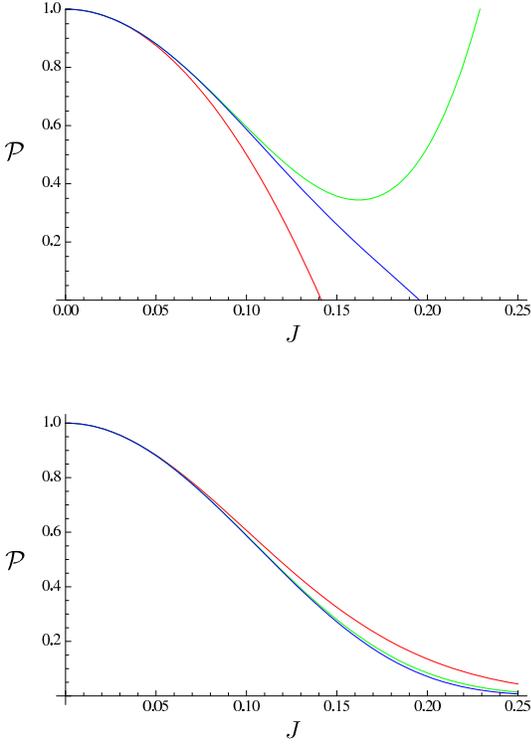}
  \caption{(color online).
    ${\mathcal P}$  as a function of $J_x=J_y=J$ computed for $m=25$. Top :
    nonresummed (bare) expression (\ref{eq:proba}) at order 2 [red (bottom)], 4 [green (top)] and 6 [blue (middle)]. Bottom : exponentiated form $\exp(A- m B)$ at order 2 [red (top)], 4 [green (middle)] and 6 [blue (bottom)].}
  \label{fig:string_proba}
\end{figure}

We display the results at various order in Fig.~\ref{fig:string_proba} using the expanded form (\ref{eq:proba}) and the exponential form. As can be clearly seen, the exponential form seems  to be well-behaved. In addition, the order 6 expansion of $A$ and $B$ seems to provide an almost converged result when put in the exponential. 
Thus, we claim that one can use this form to obtain a very accurate value of ${\mathcal P}$ which is known to be of primer interest for braiding experiments \cite{Dusuel08_1,Jiang08,Aguado08,Nori08}. 
%
%
%
%

%
%
\subsection{Anyons without fermions}
\label{subsec:awf}
%
%
As discussed previously, local or string spin operations create fermions. However, in experiments, one wishes to manipulate anyons without being spoiled by these fermions \cite{Jiang08}. In other words, the ideal operations would consist in exciting plaquettes (only) while remaining in the ground state of the corresponding vortex sector. In this section, we shall show, perturbatively,  that it is possible to do so even if the form of such operators is hard to implement in realistic devices.

As an example, let us determine the operator $\Omega_\bi$ creating two vortices at the left and right plaquettes of a given site $\bi$ [see inset Fig.~\ref{fig:Iyz} (left)]. 
This operator must be such that ${\Omega_\bi}_\mathrm{eff}=U^\dagger \Omega_\bi U=\tau_\bi^z$ which indeed leads to $I_0^z=1$. Note that this procedure is the inverse of what is usually done with CUTs since, here, we wish to compute the bare observable given the effective observable instead of the opposite.

Let us assume that this operator has a perturbative expansion, namely
%
%
\begin{equation}
\Omega_\bi=\sum_{k\in \mathbb{N}} \Omega_\bi^{(k)},
\end{equation}
%
%
where $\Omega_\bi^{(k)}$ contains all operators of order $k$ and thus associated to $J_x^l J_y^m$ 
(with $l+m=k$).
At order 0, operators are not renormalized so that one obviously has $\Omega_\bi^{(0)}=\tau_\bi^z$. 
The renormalization of $\Omega_\bi$ under the unitary transformation $U$ reads
%
%
\begin{eqnarray}
  \label{eq:expansion_Omega_eff}
  {\Omega_\bi}_\mathrm{eff}=\sum_{k\in \mathbb{N}}
  {\Omega_\bi}_\mathrm{eff}^{(k)}
  =\sum_{k\in \mathbb{N}} U^\dag \Omega_\bi^{(k)} U
  =\sum_{k \in \mathbb{N}}\sum_{l\in \mathbb{N}}
  {{\Omega_\bi}_\mathrm{eff}^{(k),[l]}},
\end{eqnarray}
%
%
where ${{\Omega_\bi}_\mathrm{eff}^{(k),[l]}}$ is of order $(k+l)$. 
Since, at order 0, one has ${\Omega_\bi}_\mathrm{eff}=\tau_\bi^z$, one must have, at each order $r>0$
%
%
\begin{equation}
\sum_{k \in \mathbb{N}}\sum_{l\in \mathbb{N}} {{\Omega_\bi}_\mathrm{eff}^{(k),[l]}}=0
\end{equation}
%
%
where the sum is restricted to values of indices such that $k+l=r$.
At order 1, this leads to 
%
%
\begin{equation}  
{{\Omega_\bi}_\mathrm{eff}^{(0),[1]}} +{{\Omega_\bi}_\mathrm{eff}^{(1),[0]}}=0. 
\end{equation}
%
%
Using Eq.~(\ref{eq:rensz1}) and the fact that ${{\Omega_\bi}_\mathrm{eff}^{(k),[0]}}=\Omega_\bi^{(k)}$, one then obtains
%
%
\begin{equation}
  \label{eq:expansion_Omega1}
  \Omega_\bi^{(1)}=-\left(J_x v_{\bi-\bn_1}^{\bi} +J_y v_{\bi-\bn_2}^{\bi}+\mbox{h.c.}\right).
\end{equation}
%
%
Using the inverse mapping of Eq.~(\ref{eq:mapping}) 
%
%
\begin{eqnarray}
\tau_\bi^x &=& \sigma_{\bi,\bullet}^x\sigma_{\bi,\circ}^x ,\\
\tau_\bi^y &=& \sigma_{\bi,\bullet}^y\sigma_{\bi,\circ}^x ,\\
\tau_\bi^z &=& \sigma_{\bi,\bullet}^z ,\\
\crb_\bi &=&\frac{1}{2}\left(\sigma_{\bi,\circ}^x-\rmi\sigma_{\bi,\bullet}^z \sigma_{\bi,\circ}^y\right) , 
\end{eqnarray}
%
%
one finally gets, in the original spin language and at order 1, 
%
%
\begin{eqnarray}
    \label{eq:nofermion1}
    \Omega_\bi&=& \Omega_\bi^{(0)}+ \Omega_\bi^{(1)} ,\\    
    &=&\sigma_{\bi,\bullet}^z+\frac{1}{2}\Big[\\
    & &
    J_x
    \Big(
    \sigma_{\bi-\bn_1,\bullet}^z\sigma_{\bi-\bn_1,\circ}^y
    \sigma_{\bi,\bullet}^y 
    -\sigma_{\bi-\bn_1,\circ}^x\sigma_{\bi,\bullet}^x\sigma_{\bi,\circ}^z
    \Big)+ \nonumber \\
    & &
    J_y
    \Big(
    \sigma_{\bi-\bn_2,\bullet}^z\sigma_{\bi-\bn_2,\circ}^x
    \sigma_{\bi,\bullet}^x 
    -\sigma_{\bi-\bn_2,\circ}^y\sigma_{\bi,\bullet}^y\sigma_{\bi,\circ}^z
     \Big) 
    \Big].\nonumber
 \end{eqnarray}
%
%
This expression shows that to create anyons without fermions, one has to build a complex superposition of operators with fine-tuned coefficients. At order 1 considered here, such states require single and triple spin-flip operations but, of course, higher-order corrections would involve higher order spin-flip processes.
Such constraints makes  creation of anyons without fermions via local operations difficult experimentally \cite{Dusuel08_1}.

%
%
\section{Conclusion and perspectives}
\label{sec:conclusion}
%
%

We have analyzed perturbatively the gapped phase of the Kitaev honeycomb model in the isolated-dimer limit using the continuous unitary tranformations. We have thus derived the low-energy effective theory up to order 10 which has been found to describe an interacting anyon system. This result has to be contrasted with the order 4 result which predicts a free anyon system \cite{Kitaev06}. We also showed that the excitations in each vortex sector obey fermionic statistics. 

In a second step, we focused on the action of local spin operators onto the ground state and we have shown that they generate both anyons and fermions. We also gave the form of the operator which creates anyons without fermions. This operator involves multi-spin operators which may be hard to implement experimentally.

Of course, several questions remain open in this model. As explained by Kitaev, there exists a gapless phase which is associated to non-Abelian anyons. The influence of a magnetic field in this phase is certainly one of the most challenging question and should reveal rich phenomena. Note that 
the effect of a magnetic field in the toric code already gives rise to a nontrivial phase diagram as recently discussed in Ref.~\onlinecite{Tupitsyn08,Vidal08_2}. 

Another interesting issue concerns the time evolution of local excitations. 
Indeed, in Sec.~\ref{subsec:manipulation_anyons}, we discussed the effect of a string operator onto the ground state but we always considered static quantities. Although experimentally, succesive spin operations may be performed on "short" time scales, it would be of primer interest to compute the spreading of fermionic excitations during the braiding processes proposed to detect anyons \cite{Jiang08}. 

\acknowledgments

We wish to thank M. Kamfor for a careful reading of the manuscript. K.P.S. acknowledges ESF and EuroHorcs for funding through EURYI.

\appendix

%
%
\section{Structure of the effective Hamiltonian}
\label{app:PCUT}
%
%

As we have seen in  Sec. \ref{sec:mapping}, when setting $J_z=1/2$, the
Hamiltonian (\ref{eq:ham}) can be written as
%
%
\begin{equation}
  H = -\frac{N}{2}+Q+T_0+T_{+2} + T_{-2},
\end{equation}
%
%
where $N$ is the number of $z$-dimers, $Q$ is the particle-number operator,
$T_0$ contains the pure hopping operators which does not change the number of
particles and, $T_{+2}$ $(T_{-2})$ creates (annihilates) pairs of particles. The
operators $T_{0,\pm 2}$ are proportional to the small parameters from which
the perturbation theory is performed.

The idea of the present approach is to transform the Hamiltonian
(\ref{eq:ham_v2}) into an effective one which conserves the particle number. Of
course, in general, this cannot be achieved exactly and, as often, one has to
perform a perturbative expansion. To achieve this goal, a very powerful tool is
the continuous unitary transformations method \cite{Wegner94}. For the problem
at hand, Knetter and Uhrig \cite{Knetter00} have developed a code which
computes the coefficients of this expansion at high orders \cite{Uhrig_HP}.
Practically, one must keep in mind that, at order 10 which is the maximum order
considered in this paper, one already has more than $10^4$ terms.
We refer the interested reader to Ref.~\onlinecite{Knetter00} for a detailed
derivation and we give below, for illustration, the results up to order 4.

%
%
\begin{table}[ht]
  \centering
  \begin{tabular}[t]{|c|c|c|c|}
    \hline
    Order & Operator $O$ & Coefficient $c$ & $q_\mathrm{min}$\\
    \hline
    \hline
    1 & $T_0$ & $1$ & 1\\
    \hline
    \hline    
    2 & $T_{-2} T_{+2}$ & $-1/2$ & 0\\
    \hline
    2 & $T_{+2} T_{-2}$ & $1/2$ & 2\\
    \hline
    \hline
    3 & $T_{-2}T_0T_{+2}$ & $1/4$ & 0\\
    \hline    
    3 & $T_0T_{-2}T_{+2}$ & $-1/8$ & 1\\
    \hline
    3 & $T_{-2}T_{+2}T_0$ & $-1/8$ & 1\\
    \hline
    3 & $T_{+2}T_{-2}T_0$ & $-1/8$ & 2\\    
    \hline
    3 & $T_0T_{+2}T_{-2}$ & $-1/8$ & 2\\
    \hline
    3 & $T_{+2}T_0T_{-2}$ & $1/4$ & 3\\
    \hline
    \hline
    4 & $T_{-2}T_{-2}T_{+2}T_{+2}$ & $-1/16$ & 0 \\
    \hline
    4 & $T_{-2}T_0T_0T_{+2}$ & $-1/8$ & 0 \\
    \hline
    4 & $T_{-2}T_{+2}T_{-2}T_{+2}$ & $1/8$ & 0 \\
    \hline
    4 & $T_0T_{-2}T_0T_{+2}$ & $1/8$ & 1 \\
    \hline
    4 & $T_0T_0T_{-2}T_{+2}$ & $-1/32$ & 1 \\
    \hline
    4 & $T_{-2}T_0T_{+2}T_0$ & $1/8$ & 1 \\
    \hline
    4 & $T_0T_{-2}T_{+2}T_0$ & $-1/16$ & 1 \\
    \hline
    4 & $T_{-2}T_{+2}T_0T_0$ & $-1/32$ & 1 \\
    \hline
    4 & $T_{+2}T_{-2}T_0T_0$ & $1/32$ & 2 \\
    \hline
    4 & $T_0T_{+2}T_{-2}T_0$ & $1/16$ & 2 \\
    \hline
    4 & $T_0T_0T_{+2}T_{-2}$ & $1/32$ & 2 \\
    \hline    
    4 & $T_{+2}T_{-2}T_{+2}T_{-2}$ & $-1/8$ & 2 \\
    \hline
    4 & $T_{+2}T_0T_{-2}T_0$ & $-1/8$ & 3 \\
    \hline
    4 & $T_0T_{+2}T_0T_{-2}$ & $-1/8$ & 3 \\
    \hline
    4 & $T_{+2}T_0T_0T_{-2}$ & $1/8$ & 3 \\
    \hline    
    4 & $T_{+2}T_{+2}T_{-2}T_{-2}$ & $1/16$ & 4 \\
    \hline
  \end{tabular}
  \caption{Operators appearing in $H_\mathrm{eff}$ with its corresponding
    coefficient up to order 4, together with the $q$-particle subspace they
    start to act on.}
  \label{tab:Heff}
\end{table}
%
%
The operators and corresponding coefficients are put in Table~\ref{tab:Heff},
together with the lowest number $q_\mathrm{min}$ of particles such that the
operator has, a priori, a nonzero action within the $q$-particle subspace for
$q\geqslant q_\mathrm{min}$. $q_\mathrm{min}$ is found by requiring that the
number of particles in the system is always positive, \textit{and} by using the fact
that $T_0$ projects out 0-particle states. Note that some terms may vanish for
more subtle reasons.
For example, the third order term $T_{-2} T_0 T_{+2}$ does not act on the 0-QP
states. Indeed,  $T_{+2}$ creates a 2-QP state~; then $T_0$ makes one of the
particle hop~; and finally $T_{-2}$ tries to annihilate two particles, but
cannot, since these are not nearest-neighbor anymore, due to the hopping.

One can then directly write the effective Hamiltonian
%
%
\begin{equation}
  H_\mathrm{eff} =-\frac{N}{2}+Q+ \sum_i c_i O_i,
\end{equation}
%
%
where $O_i$ is the $i^{\rm th}$ element of the column "operator"  of
Table~\ref{tab:Heff} and $c_i$ the associated coefficient, the order being given
by the first column.
By construction, the effective Hamiltonian conserves the particle number and
the energy states are ordered according to their quasiparticle  number,
the ground state being in the 0-QP sector. Furthermore, since
$[H_\mathrm{eff},Q]=0$, one may also rewrite the effective Hamiltonian in the
following form
%
%
\begin{eqnarray} 
  H_\mathrm{eff} &=&\sum_{q\in \mathbb{N}} H_{\mathrm{eff}}|_q ,
 \end{eqnarray}
%
%
where $H_\mathrm{eff} |_q$ denotes the projection of $H_\mathrm{eff}$ onto the
$q-$QP sector. Note that it is not the decoupling used in the CUTs
community where usually one gathers all operators which contain exactly $q$
creation and $q$ annihilation operators and thus act on the $q'$-QP sector with
$q'\geq q$.

Finally, one must analyze each sector defined by the number of quasiparticles and
determine the action of each operator $O_i$ in the corresponding subspace. This
is the nontrivial part of the job which depends on the problem under
consideration.
Let us emphasize that if each operator only starts to act in the
$q_\mathrm{min}$-QP sector, it has also, in general, a nontrivial action on the
$q$-QP sectors for $q>q_\mathrm{min}$.
%
%
\section{Perturbative expansion of observables}
\label{app:observable}
%
%

In this appendix, we give the general perturbative expansion of any observable $\Omega$ obtained with the CUTs using the quasiparticle number conserving generator. As is the case for the Kitaev model, we suppose that the Hamiltonian of the system can be casted in the following form
%
%
\begin{equation} 
 H = Q+ T_{-2} +T_0+T_{+2},
 \end{equation}
%
%
and satifies the hypothesis given after Eq.~(\ref{eq:hamCUT}). 
In this case, the flow equations obtained from the CUTs method can be solved perturbatively 
\cite{Knetter04}, and the effective observable can be written as:
%
%
\begin{equation} 
 \Omega_\mathrm{eff} = \Omega+ \sum_i c_i O_i,
 \end{equation}
%
%
where $O_i$ is the $i^{\rm th}$ element of the column "operator"  of Table~\ref{tab:omegaeff} and $c_i$ the associated coefficient, the order being given by the first column. 

%
%
\begin{table}[ht]
  \centering
  \begin{tabular}[t]{|c|c|c|}
    \hline
    Order & Operator & Coefficient \\
    \hline
    \hline
    1 & $T_{-2} \, \Omega$& $-1/2 $ \\
    \hline
    1 & $T_{+2} \, \Omega$& $1/2 $ \\
    \hline
    1 & $\Omega \, T_{-2}$& $1/2 $ \\
    \hline
    1 & $\Omega \, T_{+2}$& $-1/2 $ \\
    \hline
    \hline
    2 & $T_{-2} \,T_{-2} \, \Omega$& $1/8$ \\
    \hline
    2 & $T_{-2} \, T_0 \, \Omega$&   $1/4$ \\
    \hline
    2 & $T_{-2} \,T_{+2} \, \Omega$& $-1/8$ \\
    \hline
    2 & $T_{-2} \, \Omega \, T_{-2}$&   $-1/4$ \\
    \hline
    2 & $T_{-2} \, \Omega \, T_{+2}$&   $1/4$ \\
    \hline
    2 & $T_{0} \,T_{-2} \, \Omega$&   $-1/4$ \\
    \hline
    2 & $T_{0} \,T_{+2} \, \Omega$&   $-1/4$ \\
    \hline
    2 & $T_{+2} \,T_{-2} \, \Omega$&   $-1/8$ \\
    \hline
    2 & $T_{+2} \,T_{0} \, \Omega$&   $1/4$ \\
    \hline
    2 & $T_{+2} \,T_{+2} \, \Omega$&   $1/8$ \\
    \hline
    2 & $T_{+2} \, \Omega \, T_{-2}$&   $1/4$ \\
    \hline
    2 & $T_{+2} \, \Omega \, T_{+2}$&   $-1/4$ \\
    \hline
    2 & $\Omega \, T_{-2} \,  T_{-2}$&   $1/8$ \\
    \hline
    2 & $\Omega \, T_{-2} \,  T_{0}$&   $-1/4$ \\
    \hline
    2 & $\Omega \, T_{-2} \,  T_{+2}$&   $-1/8$ \\
    \hline
    2 & $\Omega \, T_{0} \,  T_{-2}$&   $1/4$ \\
    \hline
    2 & $\Omega \, T_{0} \,  T_{+2}$&   $1/4$ \\
    \hline
    2 & $\Omega \, T_{+2} \,  T_{-2}$&   $-1/8$ \\
    \hline
    2 & $\Omega \, T_{+2} \,  T_{0}$&   $-1/4$ \\
    \hline
    2 & $\Omega \, T_{+2} \,  T_{+2}$&   $1/8$ \\
    \hline
  \end{tabular}
  \caption{Operators appearing in $\Omega_\mathrm{eff}$ with the corresponding coefficient up to order 2.}
  \label{tab:omegaeff}
\end{table}
%
%

At order 6 considered in this paper for the correlation functions, there are several thousands of terms to consider. Once this effective form is derived, one then has to analyze it in the quasiparticle sector of interest as done for the effective Hamiltonian.

%
%
\section{Coefficients of the perturbative expansion of the
  Hamiltonian in the 0-QP sector}
\label{app:zeroqp}
%
%

As explained in Sec.~\ref{sec:zeroqp}, the effective Hamiltonian in the 0-QP
sector schematically reads
%
%
\begin{equation} 
  H_\mathrm{eff}|_{q=0} = E_0
  -\sum_{\{p_1,\ldots,p_n\}} C_{p_1,\ldots ,p_n} W_{p_1} W_{p_2}\ldots W_{p_n}.
\end{equation}
%
%
where $\{p_1,p_2,\ldots,p_n\}$ denotes a set of $n$ plaquettes and
$W_p$ are conserved plaquette operators. The form of the effective Hamiltonian
is translationaly invariant (of course the configuration of the $w_p$'s need not
be !), so that $C_{p_1,\ldots ,p_n}$ in fact only depends on relative
coordinates of the plaquettes, and we will use (except for the one-plaquette
coefficient) the notation
$\widetilde{C}_{p_2-p_1,\ldots,p_n-p_1}=C_{p_1,\ldots ,p_n}$.
Here, we give the perturbative expansion up to order 10 of $E_0$ and the
${\widetilde C}$'s in the limiting case $J_x,J_y \ll J_z$. 
Setting $J_z=1/2$, one gets the following results.

%
%
\subsection*{Constant term}
%
%

%
%
\begin{eqnarray}
  \frac{E_0}{N}&=&-\frac{1}{2}-\frac{J_x^2+J_y^2}{2}-\frac{J_x^4+J_y^4}{8}-\frac{J_x^6+J_y^6}{8} \nonumber\\
  &&-\frac{25}{128}\left( J_x^8 + J_y^8 \right) +\frac{9}{32} J_x^4 J_y^4 \\
  &&-\frac{49}{128}\left( J_x^{10}+J_y^{10}\right) + \frac{33}{64}\left( J_x^{6}J_y^{4}+J_x^{4}J_y^{6} \right) ,\nonumber
\end{eqnarray}
%
%
where $N$ is the number of $z$-dimers. 

%
%
\subsection*{One-plaquette term}
%
%

%
%
\begin{eqnarray}
 C_p &=&\frac{1}{2} J_x^2 J_y^2+\frac{1}{4}\left( J_x^4 J_y^2 +J_x^2 J_y^4\right)\nonumber\\
     &&+\frac{5}{16}\left( J_x^6 J_y^2 + J_x^2 J_y^6 \right)+\frac{1}{4}J_x^4 J_y^4 \nonumber\\
     &&+\frac{35}{64}\left( J_x^8 J_y^2 +J_x^2 J_y^8 \right)-\frac{59}{32}\left( J_x^6 J_y^4 +J_x^4 J_y^6\right) .\nonumber
\end{eqnarray}
%
%

%
%
\subsection*{Two-plaquette terms}
%
%

%
%
\begin{eqnarray}
\widetilde{C}_{\bn_1}&=& \frac{7}{8} J_x^4 J_y^2 -\frac{15}{16}J_x^4 J_y^4+\frac{3}{4}J_x^6 J_y^2 \nonumber\\
&&+ \frac{77}{64}J_x^8 J_y^2-\frac{55}{32}J_x^6 J_y^4-\frac{297}{128}J_x^4 J_y^6 , \nonumber\\
\widetilde{C}_{\bn_2}&=& \frac{7}{8} J_x^2 J_y^4 -\frac{15}{16}J_x^4 J_y^4+\frac{3}{4}J_x^2 J_y^6 \nonumber\\
&&+ \frac{77}{64}J_x^2 J_y^8- \frac{55}{32}J_x^4 J_y^6-\frac{297}{128}J_x^6 J_y^4 , \nonumber\\
\widetilde{C}_{\bn_1+\bn_2}&=&\frac{33}{8}J_x^4 J_y^4, \nonumber\\
\widetilde{C}_{\bn_1-\bn_2}&=&-\frac{1}{4}J_x^4 J_y^4-\frac{3}{32} \left( J_x^6 J_y^4+ J_x^4 J_y^6\right), \nonumber\\ 
\widetilde{C}_{2\bn_1}&=& -\frac{143}{32} J_x^6 J_y^4, \nonumber\\
\widetilde{C}_{2\bn_2}&=& -\frac{143}{32} J_x^4 J_y^6, \nonumber\\
\widetilde{C}_{2\bn_1+\bn_2}&=& \frac{715}{64} J_x^6 J_y^4, \nonumber\\
\widetilde{C}_{2\bn_1-\bn_2}&=& \frac{55}{64} J_x^6 J_y^4, \nonumber\\
\widetilde{C}_{\bn_1+2\bn_2}&=& \frac{715}{64} J_x^4 J_y^6, \nonumber\\
\widetilde{C}_{-\bn_1+2\bn_2}&=& \frac{55}{64} J_x^4 J_y^6. \nonumber
\end{eqnarray} 
%
%
%
%
\subsection*{Three-plaquette terms}
%
%

%
\begin{eqnarray}
\widetilde{C}_{\bn_1,2\bn_1}&=& \frac{33}{16} J_x^6 J_y^2-\frac{143}{32} J_x^6 J_y^4 + \frac{143}{64} J_x^8 J_y^2, \nonumber\\
\widetilde{C}_{\bn_2,2\bn_2}&=& \frac{33}{16} J_x^2 J_y^6-\frac{143}{32} J_x^4 J_y^6 + \frac{143}{64} J_x^2 J_y^8, \nonumber\\
\widetilde{C}_{\bn_1,\bn_1+\bn_2}&=& \frac{33}{16} J_x^4 J_y^4 -\frac{143}{128}  \left( J_x^6 J_y^4  + J_x^4 J_y^6\right), \nonumber\\
\widetilde{C}_{\bn_1,-\bn_2}&=& \frac{33}{16} J_x^4 J_y^4 -\frac{143}{128}  \left( J_x^6 J_y^4  + J_x^4 J_y^6\right), \nonumber\\
\widetilde{C}_{\bn_1,\bn_1-\bn_2}&=& -\frac{9}{16} J_x^4 J_y^4-\frac{275}{128}\left( J_x^6 J_y^4+ J_x^4 J_y^6\right) , \nonumber \\
\widetilde{C}_{\bn_1,\bn_2}&=& -\frac{9}{16} J_x^4 J_y^4-\frac{275}{128}\left( J_x^6 J_y^4+ J_x^4 J_y^6\right), \nonumber
\end{eqnarray}
\begin{eqnarray}
\widetilde{C}_{\bn_1+\bn_2,2\bn_1+\bn_2}&=& \frac{715}{64} J_x^6 J_y^4, \nonumber\\
\widetilde{C}_{\bn_1+\bn_2,\bn_1+2\bn_2}&=& \frac{715}{64} J_x^4 J_y^6, \nonumber\\
\widetilde{C}_{\bn_1,2\bn_1+\bn_2}&=& \frac{715}{64} J_x^6 J_y^4, \nonumber\\
\widetilde{C}_{\bn_2,\bn_1+2\bn_2}&=& \frac{715}{64} J_x^4 J_y^6, \nonumber\\
\widetilde{C}_{\bn_1-\bn_2,2\bn_1-\bn_2}&=& -\frac{11}{16} J_x^6 J_y^4, \nonumber\\
\widetilde{C}_{\bn_1-\bn_2,\bn_1-2\bn_2}&=& -\frac{11}{16} J_x^4 J_y^6, \nonumber\\
\widetilde{C}_{\bn_1,2\bn_1-\bn_2}&=& -\frac{11}{16} J_x^6 J_y^4, \nonumber\\
\widetilde{C}_{\bn_2,\bn_1-\bn_2}&=& -\frac{11}{16} J_x^4 J_y^6. \nonumber
\end{eqnarray}
%
%

%
%
\subsection*{Four-plaquette terms}
%
%

%
%
\begin{eqnarray}
\widetilde{C}_{\bn_1,\bn_2,\bn_1+\bn_2}&=& \frac{33}{16} J_x^4 J_y^4 ,\nonumber\\
\widetilde{C}_{\bn_1,\bn_2,-\bn_1+\bn_2}&=& \frac{55}{128} J_x^6 J_y^4 ,\nonumber\\
\widetilde{C}_{\bn_1,\bn_2,\bn_1-\bn_2}&=& \frac{55}{128} J_x^4 J_y^6 ,\nonumber
 \end{eqnarray}
\begin{eqnarray}
\widetilde{C}_{\bn_1,2\bn_1,3\bn_1}&=& \frac{715}{128} J_x^8 J_y^2 ,\nonumber\\
\widetilde{C}_{\bn_2,2\bn_2,3\bn_2}&=& \frac{715}{128} J_x^2 J_y^8 ,\nonumber\\
\widetilde{C}_{\bn_1,\bn_1+\bn_2,\bn_1+2\bn_2}&=& \frac{715}{128} J_x^4 J_y^6 ,\nonumber\\
\widetilde{C}_{\bn_2,\bn_1+\bn_2,2\bn_1+\bn_2}&=& \frac{715}{128} J_x^6 J_y^4 ,\nonumber\\
\widetilde{C}_{\bn_1,2\bn_1,2\bn_1+\bn_2}&=& \frac{715}{128} J_x^6 J_y^4 ,\nonumber\\
\widetilde{C}_{\bn_2,2\bn_2,\bn_1+2\bn_2}&=& \frac{715}{128} J_x^4 J_y^6 ,\nonumber\\
\widetilde{C}_{\bn_1,\bn_1+\bn_2,2\bn_1+\bn_2}&=& \frac{715}{128} J_x^6 J_y^4 ,\nonumber \\
\widetilde{C}_{\bn_2,\bn_1+\bn_2,\bn_1+2\bn_2}&=& \frac{715}{128} J_x^4 J_y^6,\nonumber
 \end{eqnarray} 
\begin{eqnarray}
\widetilde{C}_{\bn_1,2\bn_1,\bn_2}&=& -\frac{143}{128} J_x^6 J_y^4 ,\nonumber\\
\widetilde{C}_{\bn_2,2\bn_2,\bn_1}&=& -\frac{143}{128} J_x^4 J_y^6 ,\nonumber\\
\widetilde{C}_{\bn_1,2\bn_1,2\bn_1-\bn_2}&=& -\frac{143}{128} J_x^6 J_y^4 ,\nonumber\\
\widetilde{C}_{\bn_1,\bn_1-\bn_2,\bn_1-2\bn_2}&=& -\frac{143}{128} J_x^4 J_y^6 ,\nonumber\\
\widetilde{C}_{\bn_1,\bn_1+\bn_2,2\bn_1}&=& -\frac{143}{128} J_x^6 J_y^4 ,\nonumber\\
\widetilde{C}_{\bn_1,\bn_1-\bn_2,2\bn_1}&=& -\frac{143}{128} J_x^6 J_y^4 ,\nonumber\\
\widetilde{C}_{\bn_1,\bn_2,-\bn_1}&=& -\frac{143}{128} J_x^4 J_y^6 ,\nonumber\\
\widetilde{C}_{\bn_1,\bn_1+\bn_2,\bn_1-\bn_2}&=& -\frac{143}{128} J_x^4 J_y^6 . \nonumber
\end{eqnarray} 
%
%

%
%
\subsection*{Five-plaquette terms}
%
%

%
%
\begin{eqnarray}
\widetilde{C}_{\bn_1,2\bn_1,\bn_1+\bn_2,2\bn_1+\bn_2}&=& \frac{715}{128} J_x^6 J_y^4, \nonumber\\
\widetilde{C}_{\bn_1,2\bn_1,-\bn_2,\bn_1-\bn_2}&=& \frac{715}{128} J_x^6 J_y^4, \nonumber\\
\widetilde{C}_{\bn_2,2\bn_2,\bn_1+\bn_2,\bn_1+2\bn_2}&=& \frac{715}{128} J_x^4 J_y^6, \nonumber\\
\widetilde{C}_{\bn_2,2\bn_2,-\bn_1,-\bn_1+\bn_2}&=& \frac{715}{128} J_x^4 J_y^6, \nonumber \\
\widetilde{C}_{\bn_1,2\bn_1,\bn_2,\bn_1+\bn_2}&=& -\frac{143}{128} J_x^6 J_y^4, \nonumber\\
\widetilde{C}_{\bn_1,2\bn_1,\bn_1-\bn_2,2\bn_1-\bn_2}&=& -\frac{143}{128} J_x^6 J_y^4, \nonumber\\
\widetilde{C}_{\bn_2,2\bn_2,\bn_1,\bn_1+\bn_2}&=& -\frac{143}{128} J_x^4 J_y^6, \nonumber\\
\widetilde{C}_{\bn_2,2\bn_2,-\bn_1+\bn_2,-\bn_1+2\bn_2}&=& -\frac{143}{128} J_x^4 J_y^6. \nonumber
\end{eqnarray} 
%

%
%
\subsection*{Six-plaquette terms}
%
%

%
%
\begin{eqnarray}
\widetilde{C}_{\bn_1,2\bn_1,\bn_2,\bn_1+\bn_2,2\bn_1+\bn_2}&=& \frac{715}{128} J_x^6 J_y^4, \nonumber\\
\widetilde{C}_{\bn_2,2\bn_2,\bn_1,\bn_1+\bn_2,\bn_1+2\bn_2}&=& \frac{715}{128} J_x^4 J_y^6 . \nonumber
\end{eqnarray} 
%

As can be seen from these expansions, the number of interacting plaquettes increases with the order of the perturbation theory.    
%
%
\section{Coefficients of the perturbative expansion of the Hamiltonian in the 1-QP sector}
\label{app:oneqp}
%
%

In the 1-QP sector, the effective Hamiltonian reads [see Eqs.~(\ref{eq:ham1qp})-(\ref{eq:S_and_t})]
%
%
\begin{equation}
  H_\mathrm{eff}|_{q=1} = H_\mathrm{eff}|_{q=0}+ \mu 
  -\sum_{\{\bj_1,\dots,\bj_n\}} D_{\bj_1,\dots ,\bj_n}
  t^{\bj_n}_{\bj_{n-1}}\ldots t^{\bj_2}_{\bj_1},
\end{equation}
%
%
where the sum is performed over all non self-retracing paths of length $n$
starting at site $\bj_1$ and ending at site $\bj_n$. The operators $t_\bi^\bj$ are defined in 
Eqs.~(\ref{eq:defQ}-\ref{eq:deftv}).
Since the $D$'s do not depend on the initial site, we introduce
${\widetilde D}_{\bj_2-\bj_1,\,\ldots,\bj_n-\bj_{n-1}}
=D_{\bj_1,\ldots,\bj_n}$. From the symmetries of the underlying lattice, it is clear that we can limit the analysis to processes involving a first jump in the direction $+\bn_1$ or $+\bn_2$. 


We give below the perturbative expansion of $\mu$ and the ${\widetilde D}$'s in the limiting case $J_x,J_y \ll J_z$ up to order 4 and set $J_z=1/2$. Note that one could reach order 10 as for the 0-QP sector if needed. However, as explained in Sec.~\ref{sec:oneqp}, it is simpler, in this sector, to use directly the Majorana formalism which is nonperturbative and requires a comparable numerical effort.

%
%
\subsection*{Chemical potential}
%
%

%
%
\begin{equation}
\mu =1+ J_x^2+J_y^2+\frac{J_x^4+J_y^4}{4}.
\end{equation}
%
%

%
%
\subsection*{One-hopping terms}
%
%

%
%
\begin{eqnarray}
 {\widetilde D}_{n_1} &=& J_x-\frac{1}{2} J_x^3-\frac{1}{2} J_x J_y^2 ,\nonumber\\
 {\widetilde D}_{n_2} &=& J_y-\frac{1}{2} J_y^3-\frac{1}{2} J_x^2 J_y. \nonumber
\end{eqnarray}
%
%

%
%
\subsection*{Two-hopping terms}
%
%

%
%
\begin{eqnarray}
 {\widetilde D}_{n_1,n_1} &=& \frac{1}{2} J_x^2-\frac{5}{8} J_x^2 J_y^2-\frac{1}{2} J_x^4 ,\nonumber\\
 {\widetilde D}_{n_1,n_2} &=& \frac{1}{2} J_x J_y-\frac{9}{16} J_x^3 J_y-\frac{9}{16} J_x J_y^3 ,\nonumber\\
 {\widetilde D}_{n_1,-n_2} &=& -\frac{1}{2} J_x J_y ,\nonumber\\
 {\widetilde D}_{n_2,n_2} &=& \frac{1}{2} J_y^2-\frac{5}{8} J_x^2 J_y^2-\frac{1}{2} J_y^4 ,\nonumber\\
 {\widetilde D}_{n_2,n_1} &=& \frac{1}{2} J_x J_y-\frac{9}{16} J_x J_y^3-\frac{9}{16} J_x^3 J_y ,\nonumber\\
 {\widetilde D}_{n_2,-n_1} &=& -\frac{1}{2} J_x J_y.\nonumber
\end{eqnarray}
%
%

%
%
\subsection*{Three-hopping terms}
%
%

%
%
\begin{eqnarray}
 {\widetilde D}_{n_1,n_1,n_1} &=& \frac{1}{2} J_x^3,\nonumber\\
 {\widetilde D}_{n_1,n_1,n_2} &=& \frac{1}{2} J_x^2 J_y,\nonumber\\
 {\widetilde D}_{n_1,n_1,-n_2} &=& -\frac{1}{4} J_x^2 J_y,\nonumber\\
 {\widetilde D}_{n_1,n_2,n_1} &=& \frac{1}{2} J_x^2 J_y,\nonumber\\
 {\widetilde D}_{n_1,n_2,n_2} &=& \frac{1}{2} J_x J_y^2,\nonumber\\
 {\widetilde D}_{n_1,n_2,-n_1} &=& -\frac{1}{4} J_x^2 J_y,\nonumber\\
 {\widetilde D}_{n_1,-n_2,n_1} &=& 0,\nonumber\\
 {\widetilde D}_{n_1,-n_2,-n_2} &=& -\frac{1}{4} J_x J_y^2,\nonumber\\
 {\widetilde D}_{n_1,-n_2,-n_1} &=& -\frac{1}{4} J_x^2 J_y,\nonumber
\end{eqnarray}
\begin{eqnarray}
 {\widetilde D}_{n_2,n_2,n_2} &=& \frac{1}{2} J_y^3,\nonumber\\
 {\widetilde D}_{n_2,n_2,n_1} &=& \frac{1}{2} J_x J_y^2,\nonumber\\
 {\widetilde D}_{n_2,n_2,-n_1} &=& -\frac{1}{4} J_x J_y^2,\nonumber\\
 {\widetilde D}_{n_2,n_1,n_2} &=& \frac{1}{2} J_x J_y^2,\nonumber\\
 {\widetilde D}_{n_2,n_1,n_1} &=& \frac{1}{2} J_x^2 J_y,\nonumber\\
 {\widetilde D}_{n_2,n_1,-n_2} &=& -\frac{1}{4} J_x J_y^2,\nonumber\\
 {\widetilde D}_{n_2,-n_1,n_2} &=& 0,\nonumber\\
 {\widetilde D}_{n_2,-n_1,-n_1} &=& -\frac{1}{4} J_x^2 J_y,\nonumber\\
 {\widetilde D}_{n_2,-n_1,-n_2} &=& -\frac{1}{4} J_x J_y^2.\nonumber
\end{eqnarray}
%
%

%
%
\subsection*{Four-hopping terms}
%
%

%
%
\begin{eqnarray}
{\widetilde D}_{n_1,n_1,n_1,n_1} &=& \frac{5}{8} J_x^4, \nonumber\\
{\widetilde D}_{n_1,n_1,n_1,n_2} &=& \frac{5}{8} J_x^3 J_y, \nonumber\\
{\widetilde D}_{n_1,n_1,n_1,-n_2} &=& -\frac{3}{16} J_x^3 J_y, \nonumber\\
{\widetilde D}_{n_1,n_1,n_2,n_1} &=& \frac{5}{8} J_x^3 J_y, \nonumber\\
{\widetilde D}_{n_1,n_1,n_2,n_2} &=& \frac{5}{8} J_x^2 J_y^2, \nonumber\\
{\widetilde D}_{n_1,n_1,n_2,-n_1} &=& -\frac{3}{16} J_x^3 J_y, \nonumber\\
{\widetilde D}_{n_1,n_1,-n_2,n_1} &=& -\frac{1}{16} J_x^3 J_y, \nonumber\\
{\widetilde D}_{n_1,n_1,-n_2,-n_2} &=& -\frac{1}{4} J_x^2 J_y^2, \nonumber\\
{\widetilde D}_{n_1,n_1,-n_2,-n_1} &=& -\frac{1}{4} J_x^3 J_y, \nonumber
\end{eqnarray}
\begin{eqnarray}
{\widetilde D}_{n_1,n_2,n_1,n_1} &=& \frac{5}{8} J_x^3 J_y, \nonumber\\
{\widetilde D}_{n_1,n_2,n_1,n_2} &=& \frac{5}{8} J_x^2 J_y^2, \nonumber\\
{\widetilde D}_{n_1,n_2,n_1,-n_2} &=& -\frac{3}{16} J_x^2 J_y^2, \nonumber\\
{\widetilde D}_{n_1,n_2,n_2,n_1} &=& \frac{5}{8} J_x^2 J_y^2, \nonumber\\
{\widetilde D}_{n_1,n_2,n_2,n_2} &=& \frac{5}{8} J_x J_y^3, \nonumber\\
{\widetilde D}_{n_1,n_2,n_2,-n_1} &=& -\frac{3}{16} J_x^2 J_y^2, \nonumber\\
{\widetilde D}_{n_1,n_2,-n_1,-n_1} &=& -\frac{1}{4} J_x^3 J_y, \nonumber\\
{\widetilde D}_{n_1,n_2,-n_1,n_2} &=& -\frac{1}{16} J_x^2 J_y^2, \nonumber
\end{eqnarray}
\begin{eqnarray}
{\widetilde D}_{n_1,-n_2,n_1,n_1} &=& -\frac{1}{16} J_x^3 J_y, \nonumber\\
{\widetilde D}_{n_1,-n_2,n_1,n_2} &=& -\frac{1}{16} J_x^2 J_y^2, \nonumber\\
{\widetilde D}_{n_1,-n_2,n_1,-n_2} &=& \frac{1}{8} J_x^2 J_y^2, \nonumber\\
{\widetilde D}_{n_1,-n_2,-n_2,n_1} &=& 0, \nonumber\\
{\widetilde D}_{n_1,-n_2,-n_2,-n_2} &=& -\frac{3}{16} J_x J_y^3, \nonumber\\
{\widetilde D}_{n_1,-n_2,-n_2,-n_1} &=& -\frac{3}{16} J_x^2 J_y^2, \nonumber\\
{\widetilde D}_{n_1,-n_2,-n_1,-n_1} &=& -\frac{3}{16} J_x^3 J_y, \nonumber\\
{\widetilde D}_{n_1,-n_2,-n_1,-n_2} &=& -\frac{3}{16} J_x^2 J_y^2, \nonumber
\end{eqnarray}
\begin{eqnarray}
{\widetilde D}_{n_2,n_2,n_2,n_2} &=& \frac{5}{8} J_y^4, \nonumber\\
{\widetilde D}_{n_2,n_2,n_2,n_1} &=& \frac{5}{8} J_x J_y^3, \nonumber\\
{\widetilde D}_{n_2,n_2,n_2,-n_1} &=& -\frac{3}{16} J_x J_y^3, \nonumber\\
{\widetilde D}_{n_2,n_2,n_1,n_2} &=& \frac{5}{8} J_x J_y^3, \nonumber\\
{\widetilde D}_{n_2,n_2,n_1,n_1} &=& \frac{5}{8} J_x^2 J_y^2, \nonumber\\
{\widetilde D}_{n_2,n_2,n_1,-n_2} &=& -\frac{3}{16} J_x J_y^3, \nonumber\\
{\widetilde D}_{n_2,n_2,-n_1,n_2} &=& -\frac{1}{16} J_x J_y^3, \nonumber\\
{\widetilde D}_{n_2,n_2,-n_1,-n_1} &=& -\frac{1}{4} J_x^2 J_y^2, \nonumber\\
{\widetilde D}_{n_2,n_2,-n_1,-n_2} &=& -\frac{1}{4} J_x J_y^3, \nonumber
\end{eqnarray}
\begin{eqnarray}
{\widetilde D}_{n_2,n_1,n_2,n_2} &=& \frac{5}{8} J_x J_y^3, \nonumber\\
{\widetilde D}_{n_2,n_1,n_2,n_1} &=& \frac{5}{8} J_x^2 J_y^2, \nonumber\\
{\widetilde D}_{n_2,n_1,n_2,-n_1} &=& -\frac{3}{16} J_x^2 J_y^2, \nonumber\\
{\widetilde D}_{n_2,n_1,n_1,n_2} &=& \frac{5}{8} J_x^2 J_y^2, \nonumber\\
{\widetilde D}_{n_2,n_1,n_1,n_1} &=& \frac{5}{8} J_x^3 J_y, \nonumber\\
{\widetilde D}_{n_2,n_1,n_1,-n_2} &=& -\frac{3}{16} J_x^2 J_y^2, \nonumber\\
{\widetilde D}_{n_2,n_1,-n_2,-n_2} &=& -\frac{1}{4} J_x J_y^3, \nonumber\\
{\widetilde D}_{n_2,n_1,-n_2,n_1} &=& -\frac{1}{16} J_y^2 J_y^2, \nonumber
\end{eqnarray}
\begin{eqnarray}
{\widetilde D}_{n_2,-n_1,n_2,n_2} &=& -\frac{1}{16} J_x J_y^3, \nonumber\\
{\widetilde D}_{n_2,-n_1,n_2,n_1} &=& -\frac{1}{16} J_x^2 J_y^2, \nonumber\\
{\widetilde D}_{n_2,-n_1,n_2,-n_1} &=& \frac{1}{8} J_x^2 J_y^2, \nonumber\\
{\widetilde D}_{n_2,-n_1,-n_1,n_2} &=& 0, \nonumber\\
{\widetilde D}_{n_2,-n_1,-n_1,-n_1} &=& -\frac{3}{16} J_x^3 J_y, \nonumber\\
{\widetilde D}_{n_2,-n_1,-n_1,-n_2} &=& -\frac{3}{16} J_x^2 J_y^2, \nonumber\\
{\widetilde D}_{n_2,-n_1,-n_2,-n_2} &=& -\frac{3}{16} J_x J_y^3, \nonumber\\
{\widetilde D}_{n_2,-n_1,-n_2,-n_1} &=& -\frac{3}{16} J_x^2 J_y^2. \nonumber
\end{eqnarray}  
%
%

Additionally, there are some terms corresponding to processes where the
particle hops one times around a plaquette. Note that the plaquette involved
can be covered clockwise or anti-clockwise but the product of $t_\bi^\bj$ leads
exactly to the same operator $\crb_\bi \anb_\bi W_p $. 
%
%
\begin{eqnarray}
{\widetilde D}_{n_1,-n_2,-n_1,n_2} &=& \frac{1}{4} J_x^2 J_y^2, \nonumber\\
{\widetilde D}_{n_1,n_2,-n_1,-n_2} &=& 0, \nonumber\\
{\widetilde D}_{n_2,-n_1,-n_2,n_1} &=& \frac{1}{4} J_x^2 J_y^2, \nonumber \\
{\widetilde D}_{n_2,n_1,-n_2,-n_1} &=& 0.\nonumber
\end{eqnarray} 
%
%

%
%
\section{Coefficients of the spin-spin correlation function in the 0-QP sector}
\label{app:CF}
%
%

As discussed in Sec.~\ref{sec:CF}, the spin-spin correlation functions 
$C_{\bi,\bj}^{\alpha \beta}=\sigma^\alpha_\bi \sigma^\beta_\bj$ computed on any eigenstate of $H$ is nonvanishing only if $\alpha=\beta$ and if $\bi$ and $\bj$ belong to the same dimer which is of $\alpha$ type.

We give below the perturbative expansion of these correlation functions in the 0-QP sector.

%
%
\subsection{Coefficients of  $C_{\bi}^{zz}$}
%
%
In the perturbative approach we use, note that a $z$-dimer in the honeycomb lattice becomes a single site $\bi$ in the effective square lattice.

\begin{figure}[t]
  \includegraphics[width=0.4\columnwidth]{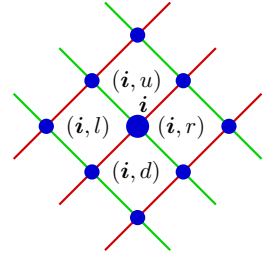}
  \caption{(color online). Labeling of the plaquettes by a site index ($\bi$)
    and a position $u,d,l,r$ with respect to that site.}
  \label{fig:observable}
\end{figure}

As for the Hamiltonian in the 0-QP sector [see Eq.~(\ref{eq:hamzeroqp})], we obtain an expansion which can be expressed only in terms of the plaquette operators, namely
%
%
\begin{equation} \label{eq:CFzzzeroqp}
C_{{\bi}}^{zz}|_{q=0} = a^{zz}-\sum_{\{p_1,\ldots,p_n\}} b^{zz}_{p_1,\ldots ,p_n} W_{p_1} W_{p_2}\ldots W_{p_n}.
\end{equation}
%
%
Below, we give the results up to order 6 and we  index a plaquette $p$ by a site $\bi$ and an indice 
$u,d,l,r$ according to notations given in Fig.~\ref{fig:observable}.

%
%
%
%

%
%
\subsubsection*{Constant term}
%
%

%
%
\begin{equation*}
a^{zz}=1-(J_x^2+J_y^2)-\frac{3}{4} (J_x^4+J_y^4)-\frac{5}{4} (J_x^6+ J_y^6).
\end{equation*}
%
%

%
%
\subsubsection*{One-plaquette terms}
%
%

%
%
\begin{eqnarray}
b_{(\bi,u)}^{zz} &=& -\frac{5}{4} J_x^2 J_y^2 ,\nonumber\\
b_{(\bi,d)}^{zz} &=& -\frac{5}{4} J_x^2 J_y^2 ,\nonumber\\
b_{(\bi,l)}^{zz} &=& -\frac{1}{4} J_x^2 J_y^2+\frac{1}{2} (J_x^2 J_y^4+ J_x^4 J_y^2), \nonumber\\
b_{(\bi,r)}^{zz} &=& -\frac{1}{4} J_x^2 J_y^2+\frac{1}{2} (J_x^2 J_y^4+ J_x^4 J_y^2), \nonumber\\
b_{(\bi+2\bn_1,l)}^{zz} &=& -\frac{21}{8} J_x^4 J_y^2, \nonumber\\
b_{(\bi+2\bn_2,r)}^{zz}  &=& -\frac{21}{8} J_x^2 J_y^4, \nonumber\\
b_{(\bi-2\bn_1,r)}^{zz} &=& -\frac{21}{8} J_x^4 J_y^2, \nonumber\\
b_{(\bi-2\bn_2,l)}^{zz} &=& -\frac{21}{8} J_x^2 J_y^4,  \nonumber\\
b_{(\bi-2\bn_1,u)}^{zz} &=& \frac{7}{8} J_x^4 J_y^2, \nonumber\\
b_{(\bi+2\bn_2,d)}^{zz} &=& \frac{7}{8} J_x^2 J_y^4, \nonumber\\           
b_{(\bi+2\bn_1,d)}^{zz} &=& \frac{7}{8} J_x^4 J_y^2, \nonumber\\
b_{(\bi-2\bn_2,u)}^{zz} &=& \frac{7}{8} J_x^2 J_y^4. \nonumber\\    
\end{eqnarray}
%
%

%
%
\subsubsection*{Two-plaquette terms}
%
%

%
%
\begin{eqnarray} 
b_{(\bi,u),(\bi+2\bn_1,l)}^{zz} &=& -\frac{21}{8} J_x^4 J_y^2, \nonumber\\ 
b_{(\bi,u),(\bi+2\bn_2,r)}^{zz} &=& -\frac{21}{8} J_x^2 J_y^4, \nonumber\\      
b_{(\bi,d),(\bi-2\bn_1,r)}^{zz}  &=& -\frac{21}{8} J_x^4 J_y^2, \nonumber\\
b_{(\bi,d),(\bi-2\bn_2,l)}^{zz}  &=& -\frac{21}{8} J_x^2 J_y^4, \nonumber\\      
b_{(\bi,l),(\bi-2\bn_1,u)}^{zz}  &=& -\frac{7}{8} J_x^4 J_y^2, \nonumber\\
b_{(\bi,l),(\bi+2\bn_2,d)}^{zz} &=& -\frac{7}{8} J_x^2 J_y^4, \nonumber\\     
b_{(\bi,r),(\bi+2\bn_1,d)}^{zz} &=& -\frac{7}{8} J_x^4 J_y^2, \nonumber\\
b_{(\bi,r),(\bi-2\bn_2,u)}^{zz}  &=& -\frac{7}{8} J_x^2 J_y^4, \nonumber     
\end{eqnarray}
\begin{eqnarray} 
b_{(\bi,u),(\bi,l)}^{zz} &=& -\frac{7}{8} J_x^4 J_y^2, \nonumber\\
b_{(\bi,d),(\bi,l)}^{zz} &=& -\frac{7}{8} J_x^2 J_y^4, \nonumber\\   
b_{(\bi,d),(\bi,r)}^{zz} &=& -\frac{7}{8} J_x^4 J_y^2, \nonumber\\
b_{(\bi,u),(\bi,r)}^{zz} &=& -\frac{7}{8} J_x^2 J_y^4. \nonumber
\end{eqnarray}
%
%

%
%
\subsection{Coefficients of $C_{\bi,\bj}^{xx}$}
%
%

Contrary to  $z$-dimers, $x$-dimers remain dimers perturbatively.  Here again, one obtains an expansion in terms of plaquettes for these observables, in the 0-QP sector, which can be written as
%
%
\begin{equation} \label{eq:CFxxzeroqp}
C_{\bi,\bi+\bn_1}^{xx}|_{q=0} = a^{xx}-\sum_{\{p_1,\ldots,p_n\}} b^{xx}_{p_1,\ldots ,p_n} W_{p_1} W_{p_2}\ldots W_{p_n},
\end{equation}
%
%

We give below the expansion of the coefficients up to order 5 (only odd orders are nonvanishing) and, as previously, we index  a plaquette $p$ by a site $\bi$ and an index $u,d,l,r$ according to the notations given in Fig.~\ref{fig:observable}. 
In the following we consider a dimer located at $(\bi,\bi+\bn_1)$. 

%
%
\subsubsection*{Constant term}
%
%

%
%
\begin{equation*}
a^{xx}=J_x +\frac{1}{2} J_x^3 +\frac{3}{4} J_x^5.
\end{equation*}
%
%

%
%
\subsubsection*{One-plaquette terms}
%
%

%
%
\begin{eqnarray}
b_{(\bi,u)}^{xx} &=&  \frac{1}{2} J_x J_y^2 - \frac{1}{2} J_x^3 J_y^2 +\frac{1}{4} J_x J_y^4, \nonumber\\
b_{(\bi,r)}^{xx}  &=&  \frac{1}{2} J_x J_y^2 - \frac{1}{2} J_x^3 J_y^2 +\frac{1}{4} J_x J_y^4, \nonumber\\
b_{(\bi+2\bn_1,l)}^{xx}  &=& \frac{7}{4} J_x^3 J_y^2, \nonumber\\
b_{(\bi,d)}^{xx} &=& \frac{7}{4} J_x^3 J_y^2, \nonumber\\
b_{(\bi+2\bn_1,d)}^{xx}  &=& -\frac{3}{4} J_x^3 J_y^2, \nonumber\\
b_{(\bi,l)}^{xx}   &=& -\frac{3}{4} J_x^3 J_y^2. \nonumber
\end{eqnarray}
%
%

%
%
\subsubsection*{Two-plaquette terms}
%
%

%
%
\begin{eqnarray}
b_{(\bi,u),(\bi+2\bn_1,l)}^{xx} &=& \frac{7}{8} J_x^3 J_y^2, \nonumber\\
b_{(\bi,u),(\bi+2\bn_2,r)}^{xx} &=& \frac{7}{8} J_x J_y^4,  \nonumber\\
b_{(\bi,u),(\bi,l)}^{xx} &=& \frac{7}{8} J_x^3 J_y^2, \nonumber\\
b_{(\bi,r),(\bi,d)}^{xx} &=& \frac{7}{8} J_x^3 J_y^2, \nonumber\\           
b_{(\bi,r),(\bi-2\bn_2,u)}^{xx} &=& \frac{7}{8} J_x J_y^4, \nonumber\\
b_{(\bi,r)(\bi+2\bn_1,d)}^{xx} &=& \frac{7}{8} J_x^3 J_y^2. \nonumber\\    
\end{eqnarray}
%
%

%
%
\subsection{Coefficients of $C_{\bi,\bj}^{yy}$}
%
%
The correlation functions $C_{\bi,\bj}^{yy}$ are straightforwardly obtained from $C_{\bi,\bj}^{xx}$ by exchanging  directions $x$ and $y$ as well as $J_x$ and $J_y$. 

%
%
\section{Correspondence between the Majorana fermion spectrum and a
  free-particle problem in a magnetic field}
\label{app:mapping}
%
%

As shown by Kitaev \cite{Kitaev06}, the spin Hamiltonian (\ref{eq:ham}) can be
mapped onto the following  Majorana fermion Hamiltonian
%
%
\begin{equation}
  H=\frac{\mathrm i}{4} \sum_{j,k} A_{j k} c_j c_k,
\end{equation}
%
%
where $A$ is a skew-symmetric matrix of size $2N \times 2N$ ($N$ being the
number of plaquette) and where the $c_j$'s are the (hermitian) Majorana
operators which obey $c_j^2=1$ and $c_j c_k=-c_k c_j$ if $j \neq k$. The sum is
performed over all sites $j$ and $k$ of the honeycomb (brickwall) lattice and
%
%
\begin{equation}
  A_{jk}=2 J_\alpha u_{jk} ,
\end{equation}
%
%
if the link $(j,k)$ is of $\alpha$-type and 0 otherwise. The $u_{jk}$'s  are
antisymmetric ($u_{jk}=-u_{kj}$) and take the values $\pm 1$. These numbers
define the vortex configuration through~: $w_p=\prod_{(j,k) \in p} u_{jk}$ where
$j$ belongs to the black sublattice and $k$ to the white one (see
Fig.~\ref{fig:mapping_brickwall_square}a).
We refer the interested reader to Ref.~\onlinecite{Kitaev06} for details.
In the very end, the whole spectrum of $H$ can be obtained once one knows the
spectrum of ${\mathrm i} A$, {\em e.~g.} the ground-state energy per plaquette is
given by
%
%
\begin{equation}
  e_0=-\frac{1}{4 N} {\rm Tr} |{\mathrm i} A|.
\end{equation}
%
%

For a bipartite lattice such as the honeycomb lattice, we shall now show that
the spectrum of ${\mathrm i} A$ is the same as the one-particle spectrum of the
following Hamiltonian
%
%
\begin{equation}
  H'=-\frac{1}{2} \sum_{j,k}  A'_{jk} (\cra_j \ana_k +{\rm h.c.}),
\end{equation}
%
%
where $\cra_j$ $(\ana_j)$ are standard spinless fermion creation (annihilation)
operators.
The Hamiltonian $H'$ describes free spinless fermions hopping in a honeycomb
lattice in a magnetic field with a flux per plaquette which equals zero
$(w_p=+1)$ or half a flux quantum ($w_p=-1$).
The spectra of $H$' (with one fermion) and of ${\mathrm i} A$ are identical
provided
%
%
\begin{equation}
  A'_{jk}=2 J_\alpha u'_{jk},
\end{equation}
%
%
with $u'_{jk}=+u'_{kj}$. The choice of the $u'_{jk}$ is as previously dictated
by the flux configuration via $w_p=\prod_{(j,k) \in p}  u'_{jk}$ where, in this
case, the $u'_{jk}$ are not oriented but still take the value $\pm 1$.

To show this, consider an eigenstate $|\psi \rangle$ of the matrix ${\rm i} A$
with energy $E$ and let us denote $\psi_j=\langle j|\psi \rangle$ its component
on site $j$. This state satisfies
%
%
\begin{equation}
  \sum_k {\rm i} A_{jk} \psi_k= E \psi_j,
\end{equation}
%
%
Since the honeycomb lattice is bipartite, we can always set $u_{jk}=u'_{jk}$ if
$j$ is a white site (and $k$ black), and $u_{jk}=-u'_{jk}$ if $j$ is a black
site (and  $k$ white). Then, one can easily check that  the state $|\phi
\rangle$ defined by $\phi_j=-\psi_j$ if $j$ is a black site and $\phi_j= -{\rm
  i}\,  \psi_j$ if it is a white site, satisfies
%
%
\begin{equation}
  -\sum_k  A'_{jk}  \phi_k= E \phi_j,
\end{equation}
%
%
so that  $|\phi\rangle$ is an eigenstate of $H'$ with the energy $E$. This shows
that $H'$ (with one particle) and ${\rm i} A$ are isospectral. We insist on the
fact that {\em this correspondence only holds for a bipartite lattice} but is no
longer true in the presence of odd cycles.


\end{document}